\documentclass[twocolumn]{aastex63}
\usepackage{xcolor}
\usepackage{amsmath}
\usepackage{float}
\usepackage{tikz}
\usetikzlibrary{shapes,arrows,snakes}
\def\ra#1#2#3{#1$^{\rm h}$#2$^{\rm m}$#3$^{\rm s}$}
\def\dec#1#2#3{#1$^\circ$#2$'$#3$''$}

\def\nod{\nodata}

\received{October 10, 2022}

\submitjournal{ApJ}

\shorttitle{Short GRB Jets Compilation}
\shortauthors{Rouco Escorial et al.}
\graphicspath{{./}{figures/}}

\begin{document}

\title{The Jet Opening Angle and Event Rate Distributions of Short Gamma-ray Bursts from Late-time X-ray Afterglows}

\correspondingauthor{Alicia Rouco Escorial}
\email{alicia.rouco.escorial@northwestern.edu}

\newcommand{\NU}{\affiliation{Center for Interdisciplinary Exploration and Research in Astrophysics (CIERA) and Department of Physics and Astronomy, Northwestern University, 1800 Sherman Ave, Evanston, IL 60201, USA}}
\newcommand{\GSFC}{\affiliation{NASA Goddard Space Flight Center, University of Maryland, Baltimore County, Greenbelt, MD 20771, USA}}
\newcommand{\CfA}{\affiliation{Center for Astrophysics\:$|$\:Harvard \& Smithsonian, 60 Garden St. Cambridge, MA 02138, USA}}
\newcommand{\Einstein}{\altaffiliation{NASA Einstein Fellow}}
\newcommand{\NASA}{\altaffiliation{NASA Postdoctoral Fellow}}
\newcommand{\UAH}{\affiliation{Center for Space Plasma and Aeronomic Research, University of Alabama in Huntsville, 320 Sparkman Drive, Huntsville, AL 35899, USA}}
\newcommand{\USRA}{\affiliation{Science and Technology Institute, Universities Space Research Association, Huntsville, AL 35805, USA}}
\newcommand{\Arizona}{\affiliation{University of Arizona, Steward Observatory, 933 N. Cherry Avenue, Tucson, AZ 85721, USA}}
\newcommand{\Bath}{\affiliation{Department of Physics, University of Bath, Claverton Down, Bath, BA2 7AY, UK}}
\newcommand{\OU}{\affiliation{Astrophysical Institute, Department of Physics and Astronomy, 251B Clippinger Lab, Ohio University, Athens, OH 45701, USA}}
\newcommand{\Adler}{\affiliation{The Adler Planetarium, Chicago, IL 60605, USA}}
\newcommand{\GeminiN}{\affiliation{Gemini Observatory/NSF's NOIRLab, 670 N. A'ohoku Place, Hilo, HI, 96720, USA}}
\newcommand{\UMD}{\affiliation{Joint Space-Science Institute, University of Maryland, College Park, MD 20742, USA}}
\newcommand{\GWU}{\affiliation{Department of Physics, The George Washington University, Washington, DC 20052, USA}}
\newcommand{\Leicester}{\affiliation{School of Physics and Astronomy, University of Leicester, University Road, Leicester, LE1 7RH, UK}}
\newcommand{\Marin}{\affiliation{College of Marin, 120 Kent Avenue, Kentfield 94904 CA, USA}}
\newcommand{\UVI}{\affiliation{University of the Virgin Islands, \#2 Brewers bay road, Charlotte Amalie, 00802 USVI, USA}}
\newcommand{\Radboud}{\affiliation{Department of Astrophysics/IMAPP, Radboud University, 6525 AJ Nijmegen, The Netherlands}}
\newcommand{\Warwick}{\affiliation{Department of Physics, University of Warwick, Coventry, CV4 7AL, UK}}
\newcommand{\Birmingham}{\affiliation{Birmingham Institute for Gravitational Wave Astronomy and School of Physics and Astronomy, University of Birmingham, Birmingham B15 2TT, UK}}
\newcommand{\Edinburgh}{\affiliation{Institute for Astronomy, University of Edinburgh, Royal Observatory, Blackford Hill, EH9 3HJ, UK}}
\newcommand{\Caltech}{\affiliation{Cahill Center for Astrophysics, California Institute of Technology, 1200 E. California Blvd. Pasadena, CA 91125, USA}}
\newcommand{\LJMU}{\affiliation{Astrophysics Research Institute, Liverpool John Moores University, 146 Brownlow Hill, Liverpool L3 5RF, UK}}
\newcommand{\CCA}{\affiliation{Center for Computational Astrophysics, Flatiron Institute, 162 W. 5th Avenue, New York, NY 10011, USA}}
\newcommand{\Columbia}{\affiliation{Department of Physics and Columbia Astrophysics Laboratory, Columbia University, New York, NY 10027, USA}}
\newcommand{\CRESST}{\affiliation{Center for Research and Exploration in Space Science and Technology (CRESST) and NASA Goddard Space Flight Center, Greenbelt, MD 20771, USA}}
\newcommand{\Maryland}{\affiliation{Department of Physics, University of Maryland, Baltimore County, 1000 Hilltop Circle, Baltimore, MD 21250, USA}}
\newcommand{\MPIA}{\affiliation{Max-Planck-Institut f\"ur Astronomie (MPIA), Königstuhl 17, 69117 Heidelberg, Germany}}
\newcommand{\Berkely}{\affiliation{Department of Astronomy, University of California, Berkeley, CA 94720-3411, USA}}
\newcommand{\IU}{\affiliation{Department of Astronomy, Indiana University, Bloomington, IN 47405-7105, USA}}
\newcommand{\Cornell}{\affiliation{Department of Astronomy, Cornell University, 404 Space Sciences Building, Ithaca, NY 14853, USA}}
\newcommand{\UChicago}{\affiliation{University of Chicago, 5801 S Ellis Ave, Chicago, IL 60637, USA}}
\newcommand{\Utah}{\affiliation{Department of Physics and Astronomy, University of Utah, James Fletcher Building, Salt Lake City, UT 84112,USA}} 
\author[0000-0003-3937-0618]{A.~Rouco~Escorial} 
\NU

\author[0000-0002-7374-935X]{W.~Fong}
\NU

\author[0000-0002-9392-9681]{E.~Berger}
\CfA

\author[0000-0003-1792-2338]{T.~Laskar}
\Radboud\Utah

\author[0000-0002-8297-2473]{R.~Margutti}
\Berkely

\author[0000-0001-9915-8147]{G.~Schroeder}
\NU

\author[0000-0002-9267-6213]{J. C.~Rastinejad}
\NU

\author[0000-0002-1533-9037]{D.~Cornish}
\UChicago

\author[0000-0003-1130-458X]{S.~Popp}
\IU

\author[0000-0002-4443-6725]{M.~Lally}
\Cornell

\author[0000-0002-2028-9329]{A. E.~Nugent}
\NU

\author[0000-0001-8340-3486]{K.~Paterson}
\MPIA

\author[0000-0002-4670-7509]{B.~D.~Metzger}
\CCA\Columbia

\author[0000-0002-7706-5668]{R.~Chornock}
\Berkely

\author[0000-0002-8297-2473]{K.~Alexander}
\NU

\author[0000-0001-7007-6295]{Y.~Cendes}
\CfA

\author[0000-0003-0307-9984]{T.~Eftekhari}
\NU

\begin{abstract}
We present a comprehensive study of 29 short gamma-ray bursts (SGRBs) observed $\approx 0.8-60$ days post-burst using \textit{Chandra} and \textit{XMM-Newton}. We provide the inferred distributions of SGRB jet opening angles and true event rates to compare against neutron star merger rates. We perform uniform analysis and modeling of their afterglows, obtaining 10 opening angle measurements and 19 lower limits. We report on two new opening angle measurements (SGRBs 050724A and 200411A) and eight updated values, obtaining a median value of $\langle \theta_{\rm j} \rangle \approx 6.1^{\circ}$ [-3.2$^{\circ}$,+9.3$^{\circ}$] (68\% confidence on the full distribution) from jet measurements alone. For the remaining events, we infer $\theta_{\rm j}\gtrsim 0.5-26^{\circ}$. We uncover a population of SGRBs with wider jets of $\theta_{\rm j} \gtrsim 10^{\circ}$ (including two measurements of $\theta_{\rm j} \gtrsim 15^{\circ}$), representing $\sim 28\%$ of our sample. Coupled with multi-wavelength afterglow information, we derive a total true energy of $\langle E_{\rm true, tot} \rangle \approx 10^{49}-10^{50}$\,erg which is consistent with MHD jet launching mechanisms. Furthermore, we determine a range for the beaming-corrected event rate of $\mathfrak{R}_{\rm true} \approx360-1800$ Gpc$^{-3}$ yr$^{-1}$, set by the inclusion of a population of wide jets on the low end, and the jet measurements alone on the high end. From a comparison with the latest merger rates, our results are consistent with the majority of SGRBs originating from binary neutron star mergers. However, our inferred rates are well above the latest neutron star-black hole merger rates, consistent with at most a small fraction of SGRBs originating from such mergers.
\end{abstract}


\section{Introduction} \label{sec:comp_intro}
Short $\gamma$-ray bursts (SGRBs) are explosions characterized by the short duration of their pulses ($T_{\rm 90}\leq2$\,s) and the hardness of their spectra \citep{Kouveliotou1993}. Monitoring of the synchrotron `afterglow' \citep[e.g.,][]{Rees1992,Meszaros1993,Paradijs2000} from radio to X-rays provides unique information about the burst energetics, environment density, and opening angles of the highly collimated, relativistic `jet' launched by the central engine \citep[e.g.,][]{Rhoads1997,Panaitescu2002,Piran2005}. The joint power of the \textit{Fermi} Gamma-ray Space Telescope \citep[\textit{Fermi};][]{GLAST1999} and the Neil Gehrels \textit{Swift} Observatory \citep[\textit{Swift};][]{Gehrels2004} has been fundamental in their discovery and arcsecond-scale localizations \citep[e.g.,][]{Lien2016,vonKienlin+20GBM10yrcat}, as well as facilitating the immediate observational follow-up of more than 100 SGRBs \citep[e.g.,][]{Evans2007,Kann2011,Margutti2013,DAvanzo2014,Fong2015a}. These studies have not only provided insightful information about their afterglows, but also enabled associations to host galaxies and subsequent redshifts  \citep[e.g.,][]{Berger2006,Wainwright2007,Berger2009,Fong2010,Church2011,Berger2014a,BRIGHT-I,BRIGHT-II, OConnor2022}.

One of the most important parameters in determining the energy scales, rates and different progenitor channels of SGRBs is the opening angle of the relativistic outflow \citep[e.g.,][]{Frail2001,Fong2015a,Mandhai2018}, which can be inferred from the afterglow light curves. The opening angle is key in the determination of the true energy scale and rates of these events \citep[e.g.,][]{Frail2001,Fong2015a,Mandhai2018}, which are fundamental properties that can help differentiate between potential progenitor channels. Considering an observer aligned with the jet axis, the jet opening angles can be calculated from the detection of temporal steepenings, or `jet breaks' in the broad-band afterglow light curves \citep{Piran1999,Rhoads1999,Sari1999,Panaitescu2005}.

In this effort, {\it Swift} has played an essential role facilitating X-ray follow-up of the SGRB afterglows \citep{Evans2007,Evans2009,Nysewander2009,Racusin2009,Margutti2013}, especially at $\lesssim1$~days after the burst trigger. However, SGRB afterglows are generally fast-fading, less energetic than long-duration GRBs, and hence are significantly fainter beyond 1 day \citep[e.g.,][]{Gehrels2008,Kann2010,Berger2013a,Fong2015a}, when jet breaks potentially take place. Therefore, dedicated deep monitoring campaigns with the sensitivity of the \textit{Chandra} X-ray observatory \citep[\textit{Chandra};][]{Weisskopf2000} and the X-ray Multi-Mirror Mission Newton \citep[\textit{XMM-Newton;}][]{Jansen2001} are necessary to detect jet breaks or place meaningful limits on their collimation. These observations are often supported by joint observations in the optical and radio bands \citep[e.g.,][]{Nicuesa2012,Fong2014,Fong2015a,Troja2016,Lamb2019,Troja2019,Fong2021}. So far, a handful of SGRB jet opening angles of $\approx 2-7^{\circ}$ have been measured \citep[e.g.][]{Burrows2006,Soderberg2006,Fong2012,Troja2016,Lamb2019,OConnor2021}, whereas for the remaining cases, lower limits of $\gtrsim 4-25^{\circ}$ have been inferred \citep[e.g.,][]{Fong2015a,Jin2018,Rouco2021}. A priori, not much is known about the general description for jet's formation and structure \citep[e.g.,][]{Blandford1977,Rosswog2002,Ruiz2016,Granot2002,Lamb2021,Margutti2021,Gottlieb2022}, nor about the true distribution of SGRB jet opening angles \citep[e.g.,][]{Sari1999,MeszarosRees2001,Zhang2003}, and therefore, nor about the real event rate distribution.

Previous studies have shown that SGRBs release a kinetic energies of $\approx 10^{49}$\,erg, and explode in low-density environments, i.e. $\approx 10^{-3}$--$10^{-2}$\,cm$^{-3}$ \citep{Nakar2007,Nicuesa2012,Berger2014a,Fong2015a}, commensurate to the significant offsets from their host galaxies \citep[][]{FongBerger2013,OConnor2022,BRIGHT-I}. The discovery of the first binary neutron star (BNS) merger gravitational-wave event, GW170817 \citep{Abbott2017a}, in conjunction with SGRB~170817A \citep[e.g.][]{Abbott2017a,Goldstein2017,Savchenko2017}, and the kilonova AT2017gfo \citep[][and references therein]{Margutti2021}, provided the first direct evidence that at least some SGRBs originate from BNS mergers. In addition, the jet of GW170817 was observed off-axis \citep[e.g.,][]{Lamb2017,Alexander2018,DAvanzo2018,Margutti2018,Troja2018,Xie2018,Hajela2019}, establishing a precedent on the study and comparison of structured jets properties against those of cosmological SGRBs, which are viewed on-axis. If observed on-axis, the GRB~170817A properties are consistent with those of cosmological SGRBs \citep[e.g.,][]{Salafia2019,Wu2019}, but required a structured outflow \citep{Margutti2021}. On the other hand, the first gravitational waves produced by neutron star-black hole (NS-BH) mergers (GW200105 and GW200115) were detected during the O3 run of Advanced LIGO/Virgo in 2021 \citep{Abbott2021}; however no electromagnetic counterparts to either detection were identified \citep[e.g.,][]{Antier2020,Coughlin2020,Anand2021,Dichiara2021,Paterson2021,Rastinejad2022}. Against this backdrop, the true fractions of BNS and NS-BH mergers which launch successful relativistic SGRB-like jets is still uncertain. 

In the coming years, the continued upgrades of gravitational wave detectors (e.g., Advanced LIGO/Virgo/KAGRA \citealt{Abbott2020b}) will refine the localization of gravitational wave events and significantly increase the number of compact binary merger detections. One can use a combination of the SGRB event rate, the BNS/NS-BH merger rates constrained by the detection of gravitational waves, and the Galactic compact-binary merger rates, to place constraints on the fraction of BNS and NS-BH mergers that produce SGRBs.  

Making use of all the available X-ray information collected by \textit{Chandra} and \textit{XMM-Newton} for \textit{Swift} SGRBs detected since 2005, we perform a comprehensive study of bursts. Utilizing the broadband afterglow information of these events available in the literature, we constrain their energetics, circumburst densities and jet opening angles. We use the distribution of these opening angles to derive the true event rate of SGRBs and compare it with the observed and theoretically-predicted NS merger rates. In Section~\ref{sec:comp_selection}, we introduce the SGRB sample. In Section~\ref{sec:comp_Xray_observations}, we describe the uniform reduction and analysis of the X-ray data. In Section~\ref{sec:comp_broadband_analysis}, we model the temporal afterglow behavior for each burst in our final sample of 29 SGRBs, and use the broad-band information of each event to constrain the energetics and circumburst densities. In Section~\ref{sec:opening_angles}, we present the distribution of SGRB jet opening angles, energy scales and event rates derived from our final sample. In Section~\ref{sec:discussion}, we discuss the implications of our findings on the energy scales and jet launching mechanisms, as well as compare our SGRB rate estimates with the BNS and NS-BH merger rates. In Section~\ref{sec:conclusions}, we summarize the main results. In the Appendix, we present the models used for fitting the afterglow light curves, and further explain the statistical test used to determine the best-fit model. We also present a panel with the X-ray observations of SGRB~130603B that show the X-ray afterglow of this burst and the contribution of a contaminant source.

The cosmology employed in this paper is standard, with H$_0=69.6$\,km~s$^{-1}$~Mpc$^{-1}$, $\Omega_{\rm{M}}=0.286$, $\Omega_{\rm{vac}}=0.714$ \citep{Bennett2014}. All the optical observations have been transformed to AB mag system and corrected for Galactic extinction in the direction of the burst \citep{Schlafly2011}. Unless otherwise noted, uncertainties on median values correspond to 16th and 84th percentiles on the full distribution.

\section{The Short Gamma-ray Burst Sample} \label{sec:comp_selection}
\begin{figure*}
\centering
\includegraphics[width=0.7\textwidth]{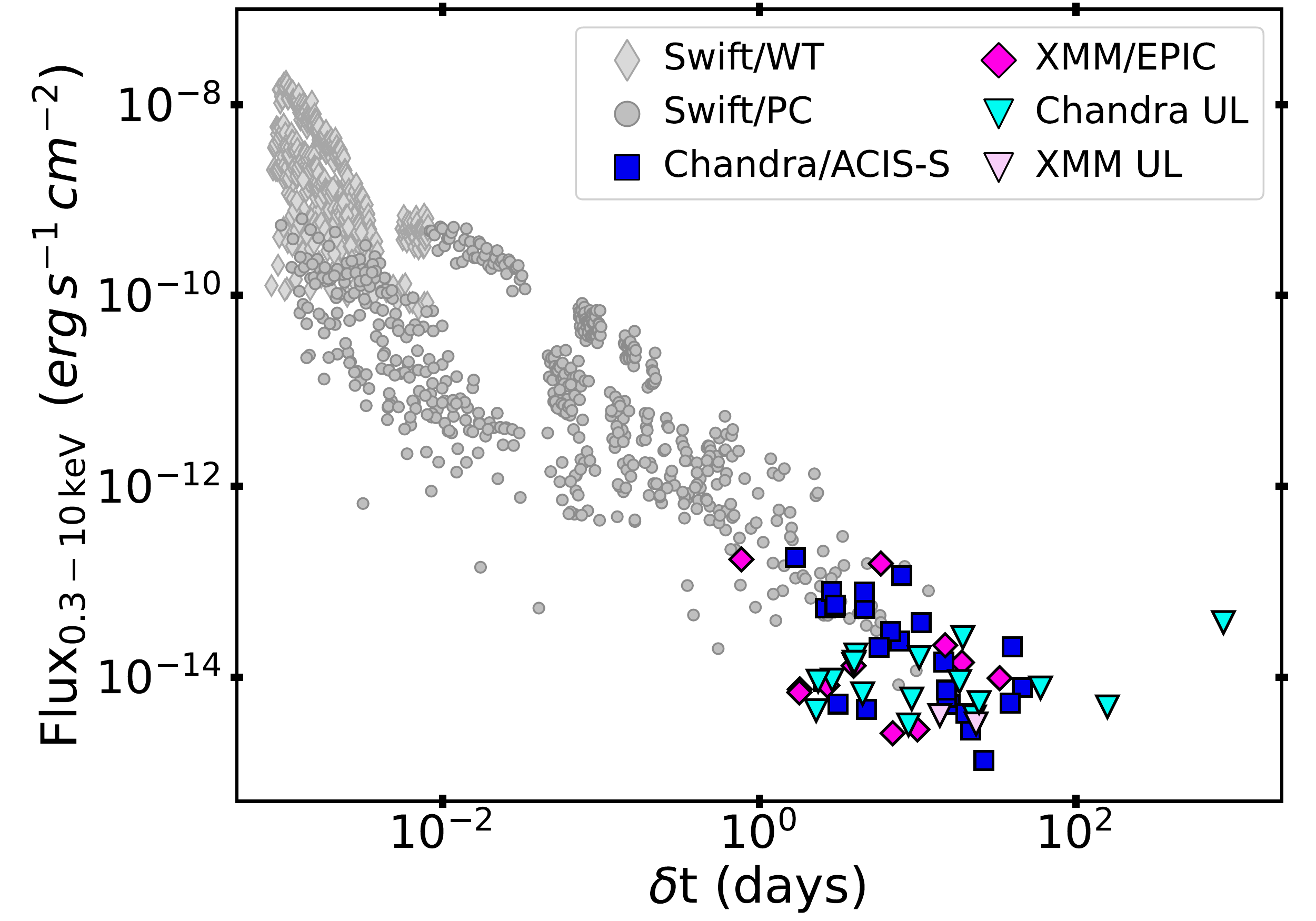}
\vspace{-0.1in}
\caption{Unabsorbed X-ray flux ($0.3-10$\,keV)  afterglow light curves of the 29 SGRBs in our sample. \textit{Swift} observations are represented as follows: WT-mode data with light grey thin diamond, and PC-mode data with dark grey circles. \textit{Chandra} and \textit{XMM-Newton} detections are depicted with dark blue squares and pink diamonds, respectively. The X-ray flux upper limits (3$\sigma$) for both observatories are shown with light blue and light pink triangles, respectively.}
\label{fig:latetime_observations}
\end{figure*}

In this paper, we present the analysis of the SGRBs that meet all the following criteria:
\begin{enumerate}
    \item \textit{Bursts classified as short, or short with extended emission (SGRB-EE)}. Generally, we select bursts with \textit{Swift}/BAT durations $T_{90}\lesssim$ 2\,s ($15$-$350$\,keV). We also include in the sample the following verified SGRB-EE events or events with potential EE \citep{Lien2016}: GRB~050724A with an initial hard, short pulse of $\sim$0.25\,s and a soft, long component longer than 100\,s in the light curve \citep{Krimm2005,Barthelmy2005b}, and GRBs~080503 \citep{Perley2009}, 101219A \citep{Krimm2010}, 150424A \citep{Barthelmy2015}, 170728B \citep{Ukwatta2017} and 200219A \citep{Laha2020}. In addition, we include GRB~050709, with $T_{90}\approx 0.07$\,s ($30-400$\,keV), that was discovered by the High Energy Transient Explorer \citep[\textit{HETE};][]{Villasenor2005,Fox2005}.
    
    \item \textit{SGRBs with \textit{Chandra} and/or \textit{XMM-Newton} observations}. In general, these observations are on the tail end or after {\it Swift}/XRT monitoring (referred as late-time observations from now on), starting at $\delta t\approx$0.8-1.7 days in the observer frame (where $\delta t$ stands for the time between the burst trigger and the mid-time of the observation; Figure\,\ref{fig:latetime_observations}).
    
    \item \textit{Events with well-sampled X-ray afterglows}. To model the X-ray afterglow and place any constraint on the decline rate, we select those events with sufficient {\it Swift}/XRT information (more than two bins in the XRT light curve) at $\delta\textnormal{t}\lesssim$1~day, that subsequently have been observed by \textit{Chandra} and \textit{XMM-Newton}. An exception for this rule is GRB~150101B, for which we ignore the XRT data since it is highly affected by the contribution of a neighboring Active Galactic Nucleus \citep{Campana2015,Fong2015b,Fong2016}. Therefore, we only use the \textit{Chandra} and \textit{XMM-Newton} data sets for this burst.
    
    \item \textit{Available broadband information of the SGRB afterglow}. We require the afterglows to have been observed in the optical and radio bands. We use detections for both bands, when not available we utilize the upper-limit information, to determine the burst energetics and environment densities that allow to obtain constraints on jet opening angles.
    
\end{enumerate}

%
\begin{longrotatetable}
\begin{deluxetable*}{lccccccccc}
\tabletypesize{\scriptsize}
\tablecaption{Basic Properties of Short GRBs with Late-time X-ray Observations}
\tablewidth{0pt}
\tablehead{
\colhead{GRB} & \colhead{RA} & \colhead{Dec} & \colhead{$\sigma_{\rm tot}$} & \colhead{$T_{\rm 90}$} & \colhead{$z$} & \multicolumn3c{Observatory} & \colhead{Redshift Reference}\\
\colhead{} & \colhead{(J2000)} & \colhead{(J2000)} & \colhead{($''$)} & \colhead{(s)} & \colhead{} & \colhead{\textit{Swift}} &
\colhead{\textit{XMM-Newton}} & \colhead{\textit{Chandra}} & \colhead{}
}
\startdata
050509B$^{*}$   &  \ra{12}{36}{13.7}   & \dec{+28}{59}{03.3}       & 5.4 & 0.0240$\pm$0.0089                   & 0.225$^{sp}$    & Y & Y & Y & \citet{Bloom2006}\\
050709$^{**}$   &  \ra{23}{01}{26.9}   & \dec{-38}{58}{39.6}       & 0.4 & 0.07$\pm$0.01$^{\dagger}$           & 0.160$^{sp}$    & N & N & Y & \citet{Villasenor2005,Fox2005}\\
050724A         &  \ra{16}{24}{44.3}   & \dec{-27}{32}{27.5}       & 0.6 & 98.7$\pm$8.6 (EE)                   & 0.257$^{sp}$    & Y & N & Y & \citet{Berger2005c}\\
               & Y & Y & N & \citet{Prochaska2006}\\
051221A         &  \ra{21}{54}{48.6}   & \dec{+16}{53}{27.0}       & 0.6 & 1.39$\pm$0.20                       & 0.546$^{sp}$    & Y & N & Y & \citet{Soderberg2006}\\
100117A         &  \ra{00}{45}{04.7}   & \dec{-01}{35}{42.1}       & 2.3 & 0.292$\pm$0.032                     & 0.915$^{sp}$    & Y & Y & N & \citet{Fong2011}\\
101219A         &  \ra{04}{58}{20.4}   & \dec{-02}{32}{23.2}       & 1.5 & 0.83$\pm$0.18 (EE)                  & 0.718$^{sp}$    & Y & N & Y & \citet{Fong2013}\\
110112A         &  \ra{21}{59}{43.7}   & \dec{+26}{27}{24.4}       & 1.9 & 0.52$\pm$0.15                       & \nod            & Y & N & Y & \citet{BRIGHT-I} \\
111020A         &  \ra{19}{08}{12.5}   & \dec{-38}{00}{43.0}       & 0.6 & 0.384$\pm$0.093                     & 0.5-1.5         & Y & Y & Y & \citet{Fong2012}\\
111117A         &  \ra{00}{50}{46.2}   & \dec{+23}{00}{39.8}       & 0.6 & 0.464$\pm$0.054                     & 2.211$^{sp}$    & Y & N & Y & \citet{Selsing2018}\\
120804A         &  \ra{15}{35}{47.5}   & \ra{-28}{46}{55.9}        & 0.6 & 0.808$\pm$0.083                     & 1.05$^{ph}$     & Y & Y & Y & \citet{BRIGHT-I}\\
130603B         &  \ra{11}{28}{48.4}   & \dec{+17}{04}{18.4}       & 1.8 & 0.176$\pm$0.024                     & 0.357$^{sp}$    & Y & Y & Y & \citet{Cucchiara2013}\\
140903A         &  \ra{15}{52}{03.2}   & \dec{+27}{36}{10.9}       & 0.6 & 0.296$\pm$0.034                     & 0.353$^{sp}$    & Y & N & Y & \citet{BRIGHT-I}\\
140930B         &  \ra{00}{25}{23.4}   & \dec{+24}{17}{39.2}       & 0.6 & 0.84$\pm$0.12                       & 1.465$^{sp}$    & Y & N & Y & \citet{BRIGHT-I}\\
150101B         &  \ra{12}{32}{05.0}   & \dec{-10}{56}{02.8}       & 0.6 & 0.0120$\pm$0.0089                   & 0.134$^{sp}$    & Y & Y & Y & \citet{Fong2016}\\
150423A         &  \ra{14}{46}{18.9}   & \dec{+12}{17}{00.3}       & 1.8 & 0.216$\pm$0.028                     & 1.392$^{sp,AG}$ & Y & N & Y &  \citet{BRIGHT-I} \\
150424A         &  \ra{10}{09}{13.3}   & \dec{-26}{37}{51.2}       & 1.4 & 81$\pm$17 (EE)                      & 0.3$^{sp}$      & Y & Y & N & \citet{Castro-Tirado2015,Schroeder2020}\\
150831A         &  \ra{14}{44}{05.8}   & \dec{-25}{38}{06.4}       & 1.9 & 0.92$\pm$0.12                       & 1.18$^{sp}$     & Y & N & Y & \citet{BRIGHT-I}\\
160624A         &  \ra{22}{00}{46.2}   & \dec{+29}{38}{37.8}       & 2.0 & 0.19$\pm$0.14                       & 0.484$^{sp}$    & Y & N & Y & \citet{OConnor2021,BRIGHT-I}\\
160821B         &  \ra{18}{39}{54.7}   & \dec{+62}{23}{30.4}       & 2.1 & 0.480$\pm$0.073                     & 0.162$^{sp}$    & Y & Y & N & \citet{Lamb2019,Troja2019,BRIGHT-I}\\
170728B         &  \ra{15}{51}{55.4}   & \dec{+70}{07}{20.6}       & 1.7 & 48$\pm$26 (EE)                      & 1.272$^{sp}$    & Y & Y & N & \citet{BRIGHT-I} \\
180418A         &  \ra{11}{20}{29.2}   & \dec{+24}{55}{59.2}       & 0.6 & 1.90$\pm$0.76$^{\ddagger}$          & 1.56$^{ph}$     & Y & N & Y & \citet{BRIGHT-I}\\
180727A         &  \ra{23}{06}{40.0}   & \dec{-63}{03}{07.1}       & 0.6 & 1.06$\pm$0.23                       & 1.95$^{ph}$     & Y & N & Y & \citet{BRIGHT-I} \\
191031D         &  \ra{18}{53}{09.5}   & \dec{+47}{38}{38.6}       & 2.0 & 0.288$\pm$0.047                     & 1.93$^{ph}$     & Y & Y & Y & \citet{BRIGHT-I}\\
200411A         &  \ra{03}{10}{39.4}   & \dec{-52}{19}{03.7}       & 1.4 & 0.220$\pm$0.045                     & 0.82$^{ph}$     & Y & N & Y & \citet{BRIGHT-I} \\
200522A         &  \ra{00}{22}{43.7}   & \dec{00}{16}{56.9}        & 0.7 & 0.616$\pm$0.079                     & 0.554$^{sp}$    & Y & N & Y & \citet{Fong2021} \\
201006A         &  \ra{04}{07}{34.3}   & \dec{+65}{09}{52.4}       & 2.1 & 0.49$\pm$0.09$^{\dagger\dagger}$    & \nod            & Y & N & Y & \citet{BRIGHT-I} \\
210726A         &  \ra{12}{53}{09.7}   & \dec{+19}{11}{24.5}       & 0.6 & 0.39$\pm$0.11$^{\ddagger\ddagger}$  & 0.37$^{ph}$     & Y & N & Y & \citet{BRIGHT-I} \\
210919A         &  \ra{05}{21}{01.0}   & \dec{+01}{18}{42.1}       & 4.7 & 0.16$\pm$0.03$^{\mathsection}$      & 0.242$^{sp}$    & Y & N & Y & \citet{BRIGHT-I}\\
211106A$^{***}$ &  \ra{22}{54}{20.5}   & \dec{-53}{13}{51.1}       & 0.6 & 1.7$\pm$0.1$^{\mathparagraph}$      & 0.5-1           & Y & Y & Y & \citet{Laskar2022} \\
\enddata
\tablecomments{
Column descriptions: (1) GRB name. (2)-(4) RA, Dec and positional uncertainty of the X-ray afterglow (see Section\,\ref{sec:comp_Xray_observations}). (5) The duration ($T_{90}$) information is retrieved from \url{https://swift.gsfc.nasa.gov/results/batgrbcat/}, except the $T_{90}$ values of GRBs~050709$^{\dagger}$, 180418A$^{\ddagger}$, 201006A$^{\dagger\dagger}$, 210726A$^{\ddagger\ddagger}$, 210919A$^{\mathsection}$ and GRB~211106A$^{\mathparagraph}$ that appear in \citet{Villasenor2005,Fox2005}, \citet{Rouco2021}, \citet{Barthelmy2020}, \citet{Palmer2021}, \citet{Barthelmy2021} and \citet{Laskar2022}, respectively. SGRBs with extended emission are indicated with (EE) (see Sec.\,\ref{sec:comp_selection}). (6) Redshifts ($z$), where ``sp'' denotes spectroscopic, ``sp,AG'' indicates spectroscopic afterglow and ``ph'' refers to photometric redshifts. (7)-(9) ``Y'' (yes) and ``N'' (no) indicate whether or not the burst position was observed by that observatory. (10) References used to retrieve redshift information.\\
$^{*}$ GRB~050509B was observed with \textit{XMM-Newton}, however none of the observations were usuful to retrieve afterglow information. $^{**}$\textit{HETE} source. $^{***}$SGRB detected by \textit{INTEGRAL} and confirmed by the \textit{Swift}/BAT-GUANO \citep{Tohuvavohu2021}.\\
}
\label{Table:log_GRB_BasicInformation}
\end{deluxetable*}
\end{longrotatetable}
%

Based on the initial SGRB sample published by \cite{Fong2015a}, and the SGRBs or SGRB-EE events detected by either \textit{Swift}/BAT\footnote{\url{https://swift.gsfc.nasa.gov/results/batgrbcat/}} or \textit{Fermi}/GBM\footnote{\url{https://heasarc.gsfc.nasa.gov/W3Browse/fermi/fermigbrst.html}} between $2015-2021$, we find a total number of 118 SGRBs or SGRB-EE, out of which 90 were followed by the \textit{Swift}/XRT and have clear X-ray afterglow detections\footnote{We exclude GRB\,080503 as the physical origin of the late-time X-ray rebrightening is unclear \citep{Perley2009}}. Applying our second criterion, the group of 90 SGRBs is reduced to 32 events observed by \textit{Chandra}, \textit{XMM-Newton} or both observatories at late times. We exclude GRBs~050813, 161104A and 210919A since their XRT light curves were only sampled by a single bin, and even in conjunction with the late-time observations, do not allow for reliable light curve modeling. All our criteria are ultimately met by 29 SGRBs that comprise our final sample. We collect a total of 60 late-time observations across all events in the sample: 18 bursts were observed by \textit{Chandra}, 4 by \textit{XMM-Newton}, and 7 additional bursts were monitored by both observatories. We present a uniform analysis of the late-time X-ray afterglow information of the 29 SGRBs with 10 bursts not being covered in the published, peer-reviewed literature: GRBs~150423A, 150424A, 150831A, 170728B, 180727A, 191031D, 200411A, 201006A, 210919A and 210726A. In Figure\,\ref{fig:latetime_observations}, we show that \textit{Chandra} and \textit{XMM-Newton} observations are critical in both the characterization of the X-ray behaviour of SGRBs and the identification of jet breaks in the light curves (see Section\,\ref{sec:opening_angles}) at $\delta t\gtrsim$0.8 days when \textit{Swift} sensitivity is not enough to follow up the faint X-ray afterglows.

We list the basic properties such as X-ray positions, durations and redshifts of the 29 SGRBs in our sample and their available X-ray afterglow information from each X-ray observatory in Table\,\ref{Table:log_GRB_BasicInformation}. Our sample has redshifts of $z\sim0.134-2.211$, which is representative of nearly the entire range of SGRBs \citep[e.g.,][]{BRIGHT-I,BRIGHT-II,OConnor2022}. Eighteen of the events have confirmed spectroscopic redshifts from their host galaxies while six have photometric redshifts. Two bursts have constrained redshift intervals (GRBs~111020A and 211106A), one case (GRB~150423A) has a redshift determined from its afterglow and two events have an inconclusive host galaxy association \citep{BRIGHT-I}. We include the redshift values in our subsequent analysis to put constraints on the explosion energetics and circumburst density properties (Table\,\ref{Table:energy_density}) of each burst in the sample, as well as determining their half-opening angles ($\theta_{\rm j}$; referred as opening angle hereinafter, see Section\,\ref{sec:opening_angles}).

%
%

\section{X-ray Observations \& Analysis}  \label{sec:comp_Xray_observations}
We perform a systematic reduction of all the available \textit{Chandra} and \textit{XMM-Newton} observations in the SGRB sample in order to model their X-ray spectra and obtain the unabsorbed X-ray flux afterglow light curves.

\subsection{{\it Chandra} Observations} \label{sec:chandra_obs}
First, we collect the 24 \textit{Chandra} observations out of 47 exposures that our team obtained through dedicated target of opportunity (ToO) and director discretionary time (DDT) programs (PIs: Berger, Fong, Rouco Escorial and Schroeder; Table\,\ref{Table:log_Xray_observations}) to follow up SGRB afterglows at late times. Additionally, we retrieved archival data of the other SGRBs (23 out of 47 exposures) in our sample utilizing the {\it Chandra} Search and Retrieval interface (ChaSeR\footnote{\url{https://cda.harvard.edu/chaser/}}). All \textit{Chandra} observations in this sample were obtained using the ACIS-S detector, except for GRB~110112A which was observed with the HRC. We used the CIAO 4.12 software package \citep{Fruscione2006} with the CALDB 4.9.0 calibration database files for reducing the data. We reprocessed the data using the \texttt{chandra\_repro} task to obtain new Level II event files, and filtered each observation to omit time intervals with high background flares (using threshold rates of $\geq0.5-0.6$ counts s$^{-1}$ in the $0.5-7$\,keV range for the ACIS-S detector). Then we ran the \texttt{CIAO} source detection tool, \textsc{wavdetect}, to perform a blind search for X-ray sources. We used the default wavelet scale values (2 and 4) to detect small features, and a threshold (\texttt{sigthresh}) of $10^{-6}$ for identifying a source pixel.  

Once we obtained a list of all detected X-ray sources for a given observation, we searched the \textit{Swift}/XRT enhanced afterglow positions for a spatially coincident {\it Chandra} source. In case of a positive match, we calculated the 1$\sigma$ uncertainty on the afterglow position as:
\begin{equation}
    \sigma_{\rm{tot}} = \sqrt[]{\sigma_{\rm{centroid}}^2 + \sigma_{\rm {astro}}^2}
    \label{eqn:astrometry}
\end{equation}

\noindent where $\sigma_{\rm{centroid}}$ is the afterglow centroid 1$\sigma$ uncertainty provided by \textsc{wavdetect} (typically $\approx 0.02-0.3''$), and $\sigma_{\rm {astro}}$ is the \textit{Chandra} absolute astrometric uncertainty ($1\sigma$) of $0.6''$\footnote{\url{https://cxc.harvard.edu/cal/ASPECT/celmon/}}. The positions and their uncertainties are listed in Table~\ref{Table:log_GRB_BasicInformation}.

We found that 14 SGRBs have at least one afterglow detection with {\it Chandra}\footnote{The only SGRB in our sample that \textit{Swift}/BAT did not observe was GRB~050709. Therefore, we used the High-Energy Transient Explorer (HETE) coordinates for this event provided by \citet{Fox2005} to confirm the \textit{Chandra} afterglow detection.}. For these bursts, we obtained the count rates and spectral information from afterglows using circular source regions centered on the \textit{Chandra} afterglow positions, with varied radii between $1.5''$ to $3.0''$ depending on the extension of the source. For the background count rates and spectra, we used source-free annuli centered on the source with default inner and outer radii of $30''$ and $60''$, respectively. In some cases, we needed to vary the sizes of these regions slightly to avoid neighboring sources.

For each detection, we extracted the afterglow net count rate using the \texttt{CIAO/}\textsc{dmextract} tool in the \textit{Chandra} $0.5-7$\,keV energy range. We also generated the source and background spectra, together with their ancillary response files (\texttt{arf}) and redistribution matrix files (\texttt{rmf}) using the \textsc{specextract} tool. In the case of non-detections, we converted the extracted \textit{Chandra} count rate ($0.5-7$\,keV) from $1.5''$ circular regions on the XRT positions into $3\sigma$ count rate upper limits using Poissonian confidence levels according to Table\,1 in \citet[][]{Gehrels1986}. 

In addition, we detected neighboring X-ray sources close to the afterglow positions of GRBs~130603B, 180418A, 200522A and 210726A. These sources could not be resolved in the source regions of the \textit{Swift}/XRT or \textit{XMM-Newton}/EPIC observations\footnote{Even though GRBs\,180418A and 200522A have already published data sets corrected for these effects \citep[][respectively]{Rouco2021,OConnor2021}, we re-perform this analysis for consistency.} and fall within the SGRB point spread function (PSF). To correct the SGRB afterglow light curves from the contributing X-ray flux of these contaminants, we extracted their spectra and modeled them to account for their spectral behavior in our analysis (see Section\,\ref{sec:xray_analysis}).

\subsection{{\it XMM-Newton} Observations} \label{sec:xmm_obs}
Similarly to Section\,\ref{sec:chandra_obs}, we gather our 7 \textit{XMM-Newton} observations of SGRBs from our ToO dedicated program (PI: Fong; Table\,\ref{Table:log_Xray_observations}). For the remaining 6 observations of SGRBs that our group did not obtain, we searched in the \textit{XMM-Newton} Science Archive (XSA\footnote{http://nxsa.esac.esa.int/nxsa-web/\#search}) for public observations. All EPIC observations were obtained utilizing the pn and both MOS detectors. We used the \textit{XMM-Newton} Science Analysis System \citep[\texttt{SAS}; version 18.0.0][]{Gabriel2004} for reducing and analyzing the observations. First, we produced calibrated event lists by running the \textsc{emproc} and \textsc{epproc} tasks and filtered them for any background flaring activity. We use threshold rates of $\geq0.4-0.5$ counts s$^{-1}$ in the $10-12$\,keV range for the pn detector, and threshold rates of $\geq0.25-0.3$ counts s$^{-1}$ in the $>$10 keV range for the MOS detectors to filter out intervals of high background.

Next, we performed a blind search for X-ray sources running the \textsc{edetectchain}\,\footnote{\url{https://www.cosmos.esa.int/web/xmm-newton/sas-thread-src-find}} routine. This \texttt{SAS} task performs a simultaneous search for EPIC sources on background-filtered images extracted in five energy bands ($0.2-0.5$\,keV for MOS and $0.3-0.5$\,keV for pn, and $0.5-1$\,keV, $1-2$\,keV, $2-4.5$\,keV, $4.5-12$\,keV for all detectors) and generates a final list of detected sources on a background-filtered image representing the full energy range. We then searched for detected sources in coincidence with the {\it Swift}/XRT positions; for relevant detected sources, we calculated the positional uncertainty (Equation~\ref{eqn:astrometry}), where $\sigma_{\rm centroid}^2$ is provided by \textsc{edetectchain} and $\sigma_{\rm {astro}}=1.5''$, the \textit{XMM-Newton} systematic error \citep{Rosen2016,delaCalle2021}.

We detected afterglows in at least one observation for 8 SGRBs. We extracted the count rates and spectra of the sources from $20''$ radius circular regions centered on the \textit{XMM-Newton} positions\footnote{For SGRBs with \textit{Chandra} and \textit{XMM-Newton} afterglow detections, we utilize the more precise \textit{Chandra} position} in the $0.3-10$\,keV energy band. To obtain the background count rate and spectral information for each detection, we used a source-free region with a size identical to the source region, randomly placed on the same detector quadrant as the afterglow. We verified that none of the data were affected by pile-up. We generated the \texttt{rmf} and \texttt{arf} files by running \textsc{rmfgen} and \textsc{arfgen}, respectively. When available, we jointly fit the EPIC/pn and EPIC/MOS spectra of each SGRB (see Section\,\ref{sec:xray_analysis}). If X-ray afterglows were not detected, we obtained 3$\sigma$ upper limits on the count rates ($0.3-10$\,keV) fixed at the positions of the \textit{Swift}/XRT afterglows using the same size of the {\it XMM-Newton} detection regions utilizing the \textsc{eupper} \texttt{SAS} task.\\

\subsection{{\it Swift} Observations} \label{sec:comp_swift}
For {\it Swift}/XRT data, we collected all of the available X-ray afterglow information from the \textit{Swift} light curve repository\footnote{\url{https://www.swift.ac.uk/xrt_curves/}} \citep{Evans2007,Evans2009}. We obtained the unabsorbed X-ray flux light curve by applying the relevant counts-to-unabsorbed flux conversion factors to the count-rate light curves, depending on the \textit{Swift} observing mode (Windowed Timing (WT) and Photon Counting (PC) modes). The automatic spectral fitting routine fits each burst's X-ray spectrum to a double-component absorbed power law model with one of the absorption components set to the Galactic column density in the direction of the burst \citep[$N_{\rm H, Gal}$;][]{Willingale2013}, and the other accounting for any excess in the line of sight as an intrinsic neutral hydrogen absorption column ($N_{\rm H, int}$), while the power-law component is characterized by the X-ray photon index ($\Gamma_X$).

For the four cases (GRBs~130603B, 180418A, 200522A and 210726A) in which the contaminating sources were discovered in \textit{Chandra} observations (Section\,\ref{sec:chandra_obs}), we utilized the option ‘create time-sliced spectra’ in the \textit{Swift} repository to obtain the spectra of each bin in the dynamically\footnote{For more information see: \url{https://www.swift.ac.uk/xrt_curves/docs.php\#products}} binned XRT light curves. Performing a spectral analysis in which we account for the contamination of these unrelated astrophysical sources leads to lower SGRB afterglow fluxes at the tail of the XRT monitoring. In Section\,\ref{sec:xray_analysis}, we describe the method to obtain the corrected unabsorbed X-ray flux light curves for each SGRB.

\subsection{X-ray Spectral Analysis} \label{sec:xray_analysis}
We analyzed all of the X-ray afterglow spectra of the 29 SGRBs in our sample utilizing \texttt{Xspec} \citep[12.10.1f;][]{Arnaud1996}. Before proceeding with the spectral analysis, we binned the spectra with \textsc{grppha} to guarantee at least one count per bin to avoid negative counts when accounting for the background during the analysis. For modeling the X-ray afterglow spectra in \texttt{Xspec}, we used \texttt{WILM} abundances \citep{wilms2000} with X-ray cross-sections set to \texttt{VERN} \citep{Verner1996} and statistics to W-statistics \citep[Cash-statistics for Poisson data and background;][]{Wachter1979}. We characterized the spectra with a two-component absorption power-law model (\texttt{tbabs*ztbabs*pow}) defined by $N_{\rm H, Gal}$ \citep[][]{Willingale2013}, $N_{\rm H, int}$ at the redshifts\footnote{If the distance to the SGRB was unknown, we set $z=0$.} of the SGRBs when available (see Table\,\ref{Table:log_GRB_BasicInformation}) and $\Gamma_X$.

Initially, we allowed all spectral parameters to vary as free except for $N_{\rm H,Gal}$ and $z$. However, we find that for 14 bursts, the 3$\sigma$ confidence interval of $N_{\rm H, int}$ was consistent with zero. Thus, we performed a revised spectral fit for these events setting $N_{\rm H,int}=0$ (see Table\,\ref{Table:log_Xray_observations}). For SGRBs initially monitored by XRT and with more than one detection with \textit{Chandra} or \textit{XMM-Newton}, we performed a joint spectral fit of the entire data set tying $\Gamma_X$ and $N_{\rm H, int}$ between spectra and leaving the normalization as a free parameter, which is valid because we did not find evidence for spectral evolution for any of the bursts. We determined the best-fit spectral parameters within the $0.5-7$\,keV and $0.3-10$\,keV energy ranges for the \textit{Chandra} and \textit{XMM-Newton} spectra, respectively. 

For SGRB afterglows that were detected by both {\it XMM-Newton} and {\it Chandra}, we set a common energy range of $0.5-7$\,keV, which is well-covered by both observatories, and used a \texttt{constant} multiplicative  model (\texttt{const*tbabs*ztbabs*pow}) to account for the cross-calibration between the different instruments of the observatories in our joint fitting. We set the \textit{XMM-Newton}/EPIC-PN constant value to 1 and calculated the rest of the constants (\texttt{const}$_{\rm {EPIC-MOS1}}=1.088$, \texttt{const}$_{\rm EPIC-MOS2}=1.106$, \texttt{const}$_{\rm ACIS-S3}=1.106$ and \texttt{const}$_{\rm XRT-PC}=0.965$) following Table~5 in \citet{Plucinsky2017}.

Finally, we determined the unabsorbed X-ray fluxes ($F_{\rm X}$) of the afterglows within the $0.3-10$\,keV energy range, fixing the spectral parameters to the best-fit values and using the \texttt{cflux} convolution model (\texttt{const*tbabs*ztbabs*cflux*pow}) in \texttt{Xspec}. The unabsorbed X-ray fluxes ($0.3-10$\,keV), together with best-fit parameters and their $1\sigma$ uncertainties are listed in Table~\ref{Table:log_Xray_observations}. We calculated the $3\sigma$ X-ray unabsorbed flux upper limits from the upper-limits of the count rates, using the \texttt{Heasarc} WebPIMMS tool\footnote{\url{https://heasarc.gsfc.nasa.gov/cgi-bin/Tools/w3pimms/w3pimms.pl}} with the best-fit spectral parameters available from previous afterglow detections for each case. The $3\sigma$ upper limits for the X-ray unabsorbed fluxes are listed in Table\,\ref{Table:log_Xray_observations}.

There are four cases with contaminating sources (X1) to the {\it Swift}/XRT or {\it XMM-Newton} source regions that are resolved in {\it Chandra} observations. We employed a single absorbed power-law model (\texttt{tbabs*pow}) to characterize their \textit{Chandra} spectra. Then we took into account these spectral parameters in the \textit{Swift} and \textit{XMM-Newton} spectral modeling of the SGRB afterglows and calculated the unabsorbed X-ray flux from the afterglow in the $0.3-10$\,keV energy band \citep[following the same method as described in][]{Rouco2021}. For that, we set \texttt{cflux} to account only for the afterglow (AG) spectral component of the model: \texttt{(tbabs*zbabs*cflux*pow)$_{\rm AG}$ + (tbabs*pow)$_{\rm X1}$}.

%
\begin{longrotatetable}
\begin{deluxetable*}{cccccccccc}
\linespread{1.3}
\tabletypesize{\scriptsize}
\tablecaption{Chandra and XMM-Newton Observations and Spectral Parameters of Short GRBs}
\tablewidth{0pt}
\tablehead{
\colhead{GRB} & \colhead{Mission} &\colhead{PI} & \colhead{ObsID} & \colhead{$\delta$t} & \colhead{Exposure} & \colhead{$N_{\rm H, Gal}$} & \colhead{$N_{\rm H, int}$} & \colhead{$\Gamma_{\rm X}$} & \colhead{$F_{\rm X}$} \\
\colhead{} & \colhead{} & \colhead{} & \colhead{} & \colhead{(days)} & \colhead{(ks)} & \colhead{(10$^{22}$\,cm$^{-2}$)} & \colhead{(10$^{22}$\,cm$^{-2}$)} & \colhead{} & \colhead{(erg~s$^{-1}$~cm$^{-2}$)}
    }
\startdata
050509B & CXO & Burrows  & $^{*}$5588               & 2.3   & 49.1 & 0.016    & --                          & 1.50$_{-0.40}^{+0.90}$     & $<4.5\times10^{-15}$ \\
050709  & CXO & Frail    & $^{\ddag}$5587           & 2.5   & 43.0 & 0.012    & --                          & 2.35$_{-0.31}^{+0.33}$     & 9.0$_{-1.3}^{+1.5}\times10^{-15}$ \\
        &     &          & $^{\ddag}$6354           & 16.0  & 18.0 & 0.012    & --                          & 2.35                       & 5.2$_{-1.6}^{+2.0}\times10^{-15}$ \\
050724A & CXO & Burrows  & $^{\ddag}$6293           & 2.6   & 49.3 & 0.277    & --                          & 1.52$\pm$0.15              & 4.8$\pm0.4\times10^{-14}$ \\
        &     &          & $^{\ddag}$5589           & 21.6  & 44.6 & 0.277    & --                          & 1.52                       & 2.6$_{-0.9}^{+1.1}\times10^{-15}$ \\
051221A & CXO & Burrows  & $^{\ddag}$6681           & 1.7   & 29.8 & 0.075    & 0.21$\pm$0.13               & 1.89$\pm0.14$              & 17.9$\pm 0.8\times10^{-14}$ \\
        &     &          & $^{\ddag}$7256           & 4.6   & 29.8 & 0.075    & 0.21                        & 1.89                       & 5.3$_{-0.4}^{+0.5}\times10^{-14}$  \\
        &     &          & $^{\ddag}$7257           & 15.3  & 17.9 & 0.075    & 0.21                        & 1.89                       & 6.2$_{-2.1}^{+2.6}\times10^{-15}$ \\
        &     &          & $^{\ddag}$7258           & 20.2  & 24.6 & 0.075    & 0.21                        & 1.89                       & 4.2$_{-1.3}^{+1.7}\times10^{-15}$ \\
        &     &          & $^{\ddag}$6683           & 26.3  & 48.2 & 0.075    & 0.21                        & 1.89                       & 1.3$_{-0.5}^{+0.7}\times10^{-15}$ \\
100117A & XMM & Schartel & 0560192001               & 1.8   & 37.8 & 0.029    & 0.56$_{-0.25}^{+0.29}$      & 2.70$_{-0.32}^{+0.35}$     & 6.9$_{-2.1}^{+2.3}\times10^{-15}$ \\
101219A & CXO & Berger   & $^{*}$11106              & 4.1   & 19.8 & 0.058    & 0.80$_{-0.40}^{+0.60}$      & 1.46$_{-0.24}^{+0.26}$     & $<1.7\times10^{-14}$ \\
110112A & CXO & La Palombara & $^{*}$14548          & 856   & 4.1  & 0.068    & 0.09$_{-0.09}^{+0.12}$      & 2.30$\pm 0.40$             & $<3.7\times10^{-14}$ \\
111020A & CXO & Berger   & $^{\ddag}$12543          & 3.0   & 19.7 & 0.143    & 0.90$_{-0.19}^{+0.21}$      & 2.03$_{-0.20}^{+0.21}$     & 5.4$\pm0.8\times10^{-14}$ \\
        &     &          & $^{**}$12544             & 10.2  & 19.8 & 0.143    & 0.90                        & 2.03                       & $<1.6\times10^{-14}$ \\
        & XMM & Schartel & $^{\ddag}$0658400901     & 0.8   & 28.4 & 0.143    & 0.90                        & 2.03                       & 17.1$\pm1.0\times10^{-14}$ \\
111117A & CXO & Sakamoto & 13784                    & 3.1   & 19.8 & 0.041    & --                          & 1.56$_{-1.42}^{+1.43}$     & 5.2$_{-1.9}^{+2.5}\times10^{-15}$ \\
120804A & CXO & Troja    & 15268                    & 9.5   & 19.8 & 0.158    & 2.36$_{-0.59}^{+0.65}$      & 1.75$_{-0.11}^{+0.12}$     & 2.8$\pm0.5\times10^{-14}$ \\
        & CXO & Burrows  & 14807                    & 46.0  & 59.3 & 0.158    & 2.36                        & 1.75                       & 7.8$_{-1.4}^{+1.6}\times10^{-15}$ \\
        & XMM & Schartel & 0700381001               & 19.1  & 33.0 & 0.158    & 2.36                        & 1.75                       & 1.4$\pm0.3\times10^{-14}$ \\
130603B & XMM & Fong     & $^{\ddag}$0722570301     & 2.7   & 16.6 & 0.021    & --                          & 2.51$_{-0.50}^{+0.57}$     & 8.2$_{-2.0}^{+2.2}\times10^{-15}$ \\
        &     &          & $^{\ddag}$0722570501     & 7.0   & 26.9 & 0.021    & --                          & 2.51                       & 2.6$_{-1.6}^{+1.4}\times10^{-15}$ \\
        & CXO &          & $^{**}$22400+23160       & 2449.6& 39.7 & 0.021    & --                          & 2.51                       & $<5.1\times10^{-15}$ \\
140903A & CXO & Sakamoto & 15873                    & 2.9   & 19.8 & 0.036    & 1.04$_{-0.52}^{+0.61}$      & 2.24$_{-0.44}^{+0.47}$     & 7.9$\pm0.9\times10^{-14}$ \\
        &     & Troja    & $^{**}$15986             & 18.4  & 58.7 & 0.036    & 1.04                        & 2.24                       & $<9.0\times10^{-15}$ \\
140930B & CXO & Fong     & 15807                    & 4.8   & 23.4 & 0.035    & --                          & 1.64$\pm 0.08$             & 4.6$_{-1.8}^{+2.4}\times10^{-15}$ \\
        &     &          & $^{**}$15808             & 22.9  & 34.3 & 0.035    & --                          & 1.64                       & $<3.8\times10^{-15}$ \\
150101B & CXO & Troja    & $^{\ddag}$17586          & 7.9   & 14.9 & 0.035    & --                          & 2.95$\pm0.13$              & 1.9$\pm0.2\times10^{-13}$\\
        &     & Levan    & $^{\ddag}$17594          & 39.7  & 14.9 & 0.035    & --                          & 2.95                       & 3.0$_{-0.6}^{+0.8}\times10^{-14}$\\
        & XMM & Fong     & $^{\ddag}$0748391101     & 6.1   & 44.6 & 0.035    & --                          & 2.95                       & 1.7$\pm0.1\times10^{-13}$\\
150423A & CXO & Berger   & $^{*}$17658              & 9.2   & 23.3 & 0.018    & 0.01$_{-0.01}^{+0.12}$      & 1.47$_{-0.22}^{+0.34}$     & $<5.9\times10^{-15}$\\
150424A & XMM & Fong     & $^{*}$0742250601         & 23.4  & 78.4 & 0.060    & 0.08$_{-0.05}^{+0.06}$      & 2.02$_{-0.17}^{+0.18}$     & $<3.2\times10^{-15}$\\
150831A & CXO & Berger   & $^{*}$16755              & 2.9   & 19.8 & 0.114    & 0.06$_{-0.06}^{+0.22}$      & 1.80$_{-0.40}^{+0.50}$     & $<9.3\times10^{-15}$\\
160624A & CXO & Troja    & $^{*}$18021              & 8.8   & 46.4 & 0.091    & 0.29$_{-0.21}^{+0.28}$      & 1.60$_{-0.40}^{+0.50}$     & $<3.2\times10^{-15}$\\
160821B & XMM & Tanvir   & $^{\ddag}$0784460301     & 3.9   & 27.4 & 0.058    & --                          & 2.22$_{-0.14}^{+0.15}$     & 1.3$\pm0.3\times10^{-14}$\\
        &     &          & $^{\ddag}$0784460401     & 10.0  & 28.5 & 0.058    & --                          & 2.22                       & 2.8$_{-2.2}^{+1.8}\times10^{-15}$\\
170728B & XMM & Fong     & 0782250301               & 5.9   & 2.0  & 0.032    & --                          & 1.15$_{-0.28}^{+0.32}$     & 1.5$\pm0.3\times10^{-13}$\\
180418A & CXO & Fong     & $^{\ddag}$20180          & 7.7   & 24.1 & 0.010    & --                          & 1.85$\pm0.14$              & 2.4$_{-0.4}^{+0.5}\times10^{-14}$\\
        &     &          & $^{**}$20181             & 19.3  & 9.8  & 0.010    & --                          & 2.66                       & $<2.6\times10^{-14}$\\
        &     &          & $^{\ddag}$21092          & 38.5  & 27.6 & 0.010    & --                          & 2.66                       & 5.3$_{-2.0}^{+2.7}\times10^{-15}$\\
180727A & CXO & Berger   & 20280                    & 6.8   & 19.8 & 0.021    & --                          & 1.61$_{-0.35}^{+0.36}$     & 2.8$\pm0.5\times10^{-14}$\\
191031D & CXO & Berger   & $^{*}$21302              & 2.4   & 19.8 & 0.066    & --                          & 1.93$_{-0.17}^{+0.28}$     & $<9.1\times10^{-15}$\\
        & XMM & D'Avanzo & $^{*}$0804070101         & 13.8  & 47.4 & 0.066    & --                          & 1.93                       & $<3.7\times10^{-15}$\\
200219A & CXO & Sakamoto & $^{*}$22451              & 4.5   & 19.2 & 0.019    & 0.02$\pm0.01$               & 1.96$_{-0.25}^{+0.35}$     & $<6.9\times10^{-15}$\\
200411A & CXO & Berger   & $^{*}$22427              & 3.9   & 19.8 & 0.016    & 0.13$_{-0.07}^{+0.09}$      & 1.68$_{-0.23}^{+0.22}$     & $<1.3\times10^{-14}$\\
200522A & CXO & Troja    & 22456                    & 5.7   & 14.8 & 0.030    & --                          & 1.38$_{-0.14}^{+0.20}$     & $2.1_{-0.5}^{+0.6}\times10^{-14}$\\
        &     &          & $^{**}$22457+23282       & 24.5  & 57.2 & 0.030    & --                          & 1.38                       & $<5.4\times10^{-15}$\\
201006A & CXO & Fong     & $^{*}$22401              & 4.0   & 24.7 & 0.450    & 0.80$_{-0.50}^{+0.70}$      & 2.10$_{-0.50}^{+0.60}$     & $<1.4\times10^{-14}$\\
210726A & CXO & Fong     & $^{\ddag}$23445          & 3.0   & 24.6 & 0.017    & --                          & 1.71$\pm0.11$              & 5.7$\pm0.7\times10^{-14}$\\
        & CXO &          & $^{\ddag}$25680          & 14.6  & 24.7 & 0.017    & --                          & 1.71                       & 1.4$_{-0.3}^{+0.4}\times10^{-14}$\\
        & CXO &          & $^{\ddag}$23446          & 15.2  & 24.6 & 0.017    & --                          & 1.71                       & 7.3$_{-2.4}^{+3.0}\times10^{-15}$\\
        & CXO & Berger   & $^{\ddag}$23542          & 4.6   & 19.8 & 0.017    & --                          & 1.71                       & 7.8$_{-0.9}^{+1.0}\times10^{-14}$\\
        & CXO & Schroeder& $^{\ddag}$26247          & 158.2 & 36.0 & 0.017    & --                          & 1.71                       & $<4.9\times10^{-15}$\\
211106A & CXO & Berger   & $^{\ddag}$23543          & 10.5  & 19.8 & 0.011    & 0.24$^{+0.15}_{-0.13}$      & 1.92$_{-0.30}^{+0.32}$     & 3.7$_{-0.6}^{+0.7}\times10^{-14}$\\
        & XMM & Fong     & $^{\ddag}$0862860301     & 14.9  & 20.3 & 0.011    & 0.24                        & 1.92                       & 2.2$\pm0.3\times10^{-14}$\\
        & XMM &          & $^{\ddag}$0862860401     & 33.0  & 46.7 & 0.011    & 0.24                        & 1.92                       & 1.0$\pm0.2\times10^{-14}$\\
        & CXO & Rouco Escorial & $^{**}$26249+26262 & 59.8  & 37.9 & 0.011    & 0.24                        & 1.92                       & $<7.7\times10^{-15}$\\
\enddata
\tablecomments{
Column descriptions: (1) GRB name. (2) Observatory that obtained the observation. (3)-(4) Principal investigator (PI) of the program and identification number of the observation (obsID). (5) Elapsed time between the burst trigger and the log-centered time of the observation ($\delta$t). (6) Final exposure time after filtering from background events. (7)-(8) Galactic column density, $N_{\rm{H, Gal}}$, and intrinsic absorption value, $N_{\rm{H, int}}$. The symbol ``--'' means that the contribution of $N_{\rm{H, int}}$ is negligible and is set to 0 in our fits. (9) X-ray photon index ($\Gamma_{\rm X}$). The spectral parameter $\Gamma_{\rm X}$ is calculated in the $0.5-7$\,keV energy range for \textit{Chandra} and between $0.3-10$\,keV for \textit{XMM-Newton} (when information from both observatories is available then a $0.5-7$\,keV energy range is applied). For observations with joint spectral analysis, only the first value of $\Gamma_{\rm X}$ is shown with uncertainties, while for the rest of observations just the value of $\Gamma_{\rm X}$ is stated. (10) $0.3-10$\,keV unabsorbed X-ray fluxes ($F_{\rm X}$). \\
Errors are 1$\sigma$ for measurements, whereas $F_\textnormal{X}$ upper limits are 3$\sigma$.\\
$^{*}$ Fixed spectral values to \textit{Swift} spectral parameters. $^{**}$ Fixed values to previous \textit{Chandra} or \textit{XMM-Newton} spectral parameters. $^{\dag}$ Flux calculated assuming default photon index of 2 for SGRBs. $^{\ddag}$ Tied spectral parameters to perform joint spectral fitting between X-ray observations.\\
}
\label{Table:log_Xray_observations}
\end{deluxetable*}
\end{longrotatetable}
%

%
\begin{figure}
\centering
\includegraphics[width=1.0\columnwidth]{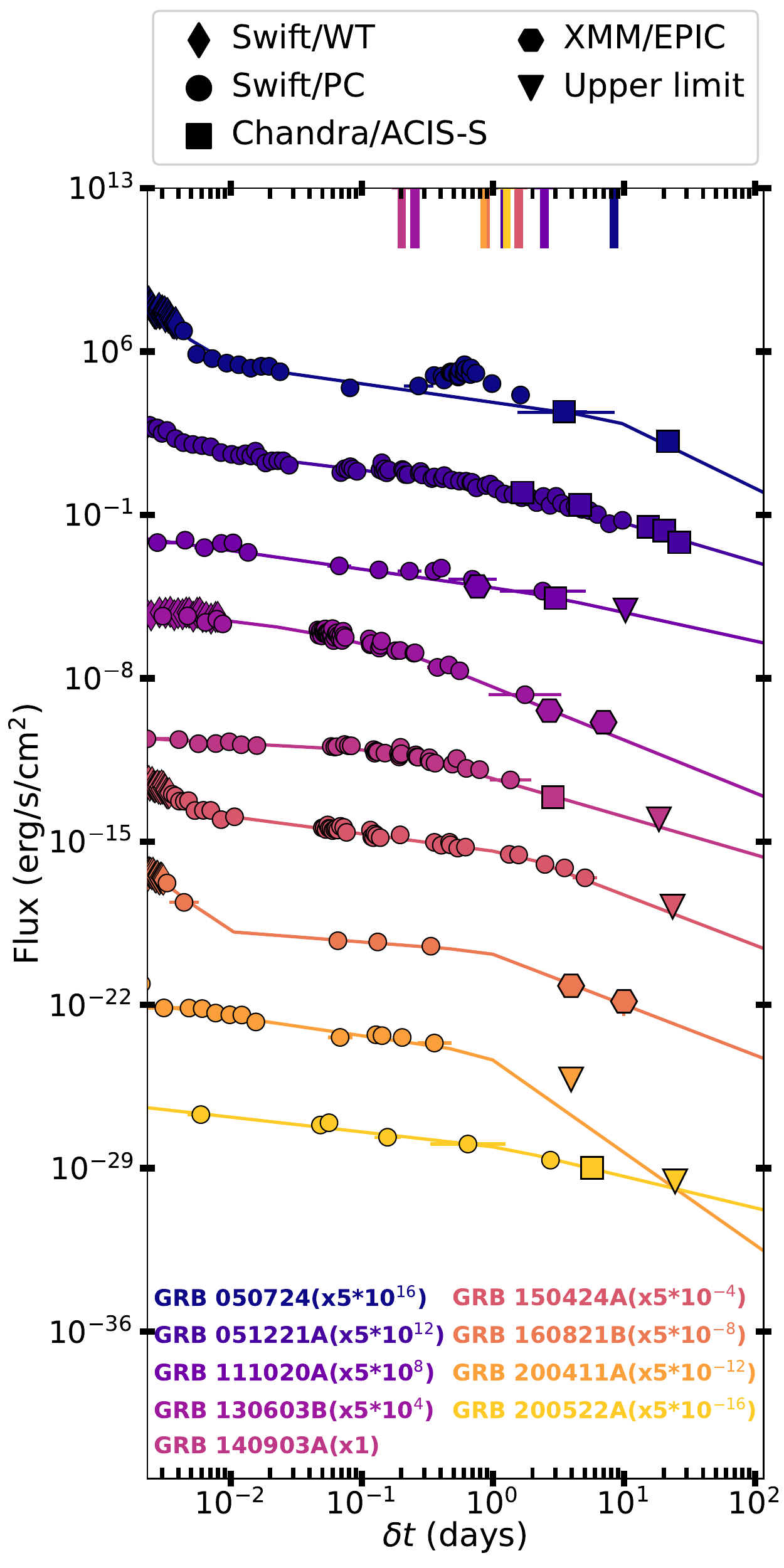}
\vspace{-0.1in}
\caption{0.3-10~keV unabsorbed X-ray flux light curves of SGRBs with detected jet breaks. All SGRBs are color-coded with their corresponding best-fit models (solid lines). Symbols indicate each set of observations obtained by different observatories or observing modes (top legend). In addition, we note that the afterglow light curve of GRB~211106A exhibits evidence for a jet in the radio band; however it is wide enough to elude detection in the X-rays. Triangles indicate 3$\sigma$ upper limits. Vertical lines from the top show the times of the jet breaks for each SGRB. Calculated uncertainties correspond to $1\sigma$.}
\label{fig:LCs_jet_breaks}
\end{figure}
%

\section{Inferred Properties from Broadband Modeling} \label{sec:comp_broadband_analysis}
Here we model the broad-band afterglows of our sample considering the standard synchrotron forward shock model \citep{Sari1998,Granot2002}. In this scenario, the forward shock originates from the interaction of a relativistic blast wave with a constant density medium, which is the expected environment for a NS merger \citep[e.g.,][]{Paczynski1986,Eichler1989}. In what follows, we first determine the temporal and spectral evolution, parameterized by $F_\nu \propto \text{t}^{\alpha} \nu^{\beta}$, where $\alpha$ and $\beta$ are the temporal and spectral power-law indices. We then use these indices and the closure relations given by the synchrotron model \citep{Granot2002} to determine the power-law index of the input distribution of accelerated electrons ($p$), providing a framework for determining the locations of the observing bands with respect to the three break frequencies of the synchrotron spectrum: the self-absorption frequency ($\nu_{sa}$), peak frequency ($\nu_{m}$) and cooling frequency ($\nu_{c}$). We then fit the available observational data using the mapping from \citet{Granot2002} to determine the burst physical properties. These properties include the isotropic-equivalent energy of the jet (E$_{\rm K, iso}$), the density of the circumburst environment ($n_0$), and the fractions of the post-shock energy transmitted to electrons ($\epsilon_{\rm e}$) and magnetic field ($\epsilon_{\rm B}$). In Section\,\ref{sec:opening_angles}, we apply these properties to determine the jet opening angles of the bursts.

\subsection{X-ray Temporal and Spectral Fitting}  \label{sec:fitting_method_LC}
The temporal behavior of X-ray afterglows can be well-described with a power-law or a series of connected power laws \citep[e.g.,][]{Zhang2006,Evans2007,Margutti2013}. To ensure that we are only using data during the forward shock afterglow phase, we ignore all of the data at $\delta t < 200$\,s, which often includes high-latitude emission and the tail of the $\gamma$-ray emission \citep[][and reference therein]{Zhang2006,Racusin2009}. Additionally, during the fits we omit portions of the light curves that are clear deviations from the main power laws\footnote{As an example, we do not include the data of GRB~050724A at $0.1< \delta t < 1.3$~days since its afterglow light curve shows a very clear X-ray flux enhancement on top of the main power-law decay trend \citep{Campana2006,Grupe2006,Margutti2011}}.

To remain agnostic to the particular best-fit model, we first fit every SGRB afterglow light curve using single, broken and triple power-law models, described by Equations~\ref{eqn:SPL}, \ref{eqn:BPL} and \ref{eqn:TPL} (see Appendix\,\ref{sec:X-ray_LC_models} for more details). The main parameters of our models are the series of temporal decay indices ($\alpha_{1,2,3}$), the flux normalization ($C$), the break times ($t_{b1,2}$), and the smoothness parameters ($s_{1,2}$). For every fit, we set all model parameters as free, except for the smoothness values which are set as constants in the broken and triple power-law models.

To perform the light curve fitting, we use the \texttt{Python} \textsc{emcee} package \citep{Foreman-Mackey2013} in which we apply an affine-invariant Markov chain Monte Carlo (MCMC) ensemble sampler \citep{Goodman2010,Foreman-Mackey2019} to find the solution with the highest log-likelihood for each model together with the model parameters. We initiate each fit with 15000 steps, each with 100 walkers, and discard the initial 500 steps as ``burn-in'' when the average log-likelihood across the chains reaches a stable value. We then calculate the medians and $1\sigma$ credible regions from the posteriors for each parameter, and use these values to determine the reduced-$\chi^2$ value for each model.

Based on the results from our light curve fitting procedure for each burst, we use the F-test to determine whether a simpler model provides a statistically better fit to our data than a more complex model. In effect, we first test the null hypothesis (H$_0$) that a single power-law model is sufficient to describe the X-ray afterglow light curve. If H$_0$ is not accepted, then we test a second null hypothesis (H'$_0$) that a broken power law is the best-fit model. If H'$_0$ is not accepted, the afterglow light curve is best-described then by a triple power law model (see Appendix~\ref{sec:F_test}). We find that twelve SGRBs are best-fit by a single power-law model, eleven by a broken power-law, and six by a triple power-law model. The medians and $1\sigma$ uncertainties of the best-fit parameters for each SGRB are shown in Table~\ref{Table:LC_fitting_parameters}. For afterglows best described by a single power-law, we find a median of $\langle\alpha_X \rangle \approx -1.00$, which is in agreement with the weighted mean of $\alpha_X$ found by \citet{Fong2015a}. 

For SGRBs with light curves best described by the broken power-law model, we find two scenarios: (i) $\alpha_1<\alpha_2$ in which the afterglow light curve transitions from a steep power-law of $\alpha_1 \lesssim -1$ to a shallower power-law of $\alpha_2 \approx [-1.5,-0.3]$, or (ii) $\alpha_1 > \alpha_2$ in which the afterglow light curve steepens. Cases in which $\alpha_2 \lesssim -2$ can be associated, in general, with jet breaks (Section\,\ref{sec:opening_angles}).

For the triple power-law model, we identify a range of morphologies in terms of afterglow behavior. Similar to the broken power-law afterglows, bursts with their final segments following the condition, $\alpha_3 \lesssim -2$ can be viable jet breaks (see Section\,\ref{sec:opening_angles} for further criteria). We note that our analysis also uncovers several instances of shallow power-law segments of $\alpha_X \approx -0.6$, which have been proposed as phases related to an extra injection of energy, for example driven by a potential magnetar central engine \citep{Rowlinson2013,Gompertz2014,Lu2017,Stratta2018}.

We also determine the X-ray spectral index ($\beta_{\rm X}$) for each SGRB. We utilize the photon indices derived from our spectral analysis in Section\,\ref{sec:comp_Xray_observations}, and the relation $\beta_{\rm X}\equiv1-\Gamma_{\rm X}$. We find that the value of $\beta_{\rm X}$\footnote{We note that the spectral parameters of SGRBs observed with \textit{Chandra} are calculated within the $0.5-7$\,keV energy band (see Section\,\ref{sec:chandra_obs}). We assume that these spectral parameters, in particular $\Gamma_{\rm X}$ and therefore $\beta_{\rm X}$, are also applicable to the $0.3-10$\,keV energy range since most of the \textit{Chandra} effective area is included in both energy ranges.} varies between -0.15 and -1.9 depending on the individual cases shown in Table\,\ref{Table:LC_fitting_parameters}. Overall we find a weighted mean of of $\langle\beta_{\rm X}\rangle = -0.99\pm 0.06$. This value is consistent within the errors with that reported by \citet{Fong2015a}.

%
\begin{deluxetable*}{cccccccc}
\tabletypesize{\footnotesize}
\linespread{1.3}
\tablecaption{SGRB Sample Light Curves Fitting Parameters}
\tablewidth{0pt}
\tablehead{
\colhead{Name} & \colhead{Model} & \colhead{$\alpha_{1}$} & \colhead{$\alpha_{2}$} & \colhead{$\alpha_{3}$} & \colhead{$\beta_{X}$} & \colhead{$t_{b1}$} & \colhead{$t_{b2}$} \\
\colhead{} & \colhead{}  & \colhead{} & \colhead{} & \colhead{} & \colhead{} & \colhead{(days)} & \colhead{(days)}
}
\startdata
\multicolumn{8}{c}{\textbf{Jet Break Detections}} \\
\hline
050724A & TPL & -4.91$_{-0.10}^{+0.12}$    & -0.80$_{-0.03}^{+0.04}$       & -2.91$_{-1.81}^{+1.15}$    & -0.52$\pm$015                & 0.0078$_{-0.0003}^{+0.0004}$          & 8.74$_{-4.85}^{+5.26}$     \\
051221A & TPL & -1.56$_{-0.09}^{+0.08}$    & -0.69$_{-3.17}^{+0.03}$       & -1.66$\pm$0.09             & -0.89$\pm$0.14               & 0.012$_{-0.002}^{+0.007}$             & 1.22$_{-0.05}^{+0.40}$     \\
111020A & BPL & -0.80$_{-0.04}^{+0.38}$    & -1.23$_{-3.22}^{+0.35}$       & --                         & -1.03$_{-0.20}^{+0.21}$      & 2.474$_{-2.465}^{+3.496}$                & --                         \\
130603B & BPL & -0.75$\pm$0.03            & -2.10$\pm$0.09               & --                         & -1.51$_{-0.50}^{+0.57}$      & 0.115$_{-0.019}^{+0.013}$                & --                         \\
140903A & TPL & -0.15$_{-3.08}^{+0.07}$    & -1.04$_{-0.25}^{+0.13}$       & -2.32$_{-1.13}^{+0.37}$    & -1.24$_{-0.44}^{+0.47}$      & 0.088$_{-0.017}^{+0.473}$             & 0.92$_{-0.42}^{+0.87}$     \\
150424A & TPL & -2.97$_{-0.23}^{+0.62}$    & -0.74$_{-2.39}^{+0.04}$       & -2.15$_{-0.69}^{+0.39}$    & -0.71$_{-0.08}^{+0.07}$      & 0.0068$_{-0.0006}^{+0.7488}$          & 1.56$_{-0.51}^{+1.45}$     \\
160821B & TPL & -4.33$_{-0.22}^{+0.23}$    & -0.44$_{-0.38}^{+0.25}$       & -2.16$_{-1.78}^{+0.65}$    & -1.22$_{-0.14}^{+0.15}$      & 0.009$\pm$0.002          & 0.89$_{-0.59}^{+1.14}$     \\
200411A & BPL & -0.82$\pm$0.05            & -3.96$_{-1.41}^{+1.62}$       & --                         & -0.72$_{-0.23}^{+0.22}$      & 0.846$_{-0.413}^{+0.916}$                & --                         \\
200522A & BPL & -0.67$\pm$0.06             & -1.74$_{-0.67}^{+0.41}$       & --                         & -0.38$_{-0.14}^{+0.20}$      & 2.410$_{-0.883}^{+1.601}$                & --                         \\
211106A$^{\dagger}$          & SPL & -0.97$\pm$0.03               & --                            & --                          & -0.92$\pm$0.30      & --                                       & --                 \\
\hline
\hline
\multicolumn{8}{c}{\textbf{Jet Break Non-Detections}} \\
\hline
050509B          & SPL & -0.91$_{-1.32}^{+0.15}$       & --                            & --                          & $^{*}$-0.50$_{-0.40}^{+0.90}$& --                                       & --                 \\
050709           & BPL & -2.75$_{-2.32}^{+1.91}$       & -0.35$_{-0.47}^{+0.22}$       & --                          & -1.35$_{-0.31}^{+0.33}$      & 2.571$_{-0.678}^{+1.778}$                   & --                 \\
100117A          & BPL & -3.42$_{-0.53}^{+0.57}$       & -1.55$_{-0.87}^{+0.22}$       & --                          & -1.70$_{-0.32}^{+0.35}$      & 0.006$_{-0.002}^{+0.004}$                & --                 \\
101219A          & SPL & -1.41$\pm 0.13$               & --                            & --                          & -0.46$_{-0.24}^{+0.26}$      & --                                       & --                 \\
110112A          & SPL & -1.04$\pm 0.05$               & --                            & --                          & -1.30$\pm 0.04$              & --                                       & --                 \\              
111117A          & SPL & -1.28$_{-0.07}^{+0.06}$       & --                            & --                          & -0.56$_{-1.42}^{+1.43}$      & --                                       & --                 \\
120804A          & SPL & -1.06$\pm 0.01$               & --                            & --                          & -0.75$_{-0.11}^{+0.12}$      & --                                       & --                 \\
140930B          & BPL & -2.07$_{-0.18}^{+0.22}$       & -1.26$_{-3.00}^{+0.06}$      & --                           & -0.64$\pm0.08$               & 0.011$_{-0.002}^{+0.008}$                & --                 \\
150101B          & SPL & -0.91$_{-0.11}^{+0.10}$       & --                            & --                          & -1.95$\pm0.13$               & --                                       & --                 \\
150423A          & SPL & -0.99$_{-0.06}^{+0.05}$       & --                            & --                          & -0.47$_{-0.22}^{+0.34}$      & --                                       & --                 \\
150831A          & TPL & -0.62$_{-0.28}^{+0.29}$       & -5.82$_{-0.13}^{+0.21}$       & -1.07$_{-0.08}^{+0.06}$     & -0.80$_{-0.40}^{+0.50}$      & 0.0018$\pm 0.0001$                       & 0.0035$\pm 0.0001$ \\
160624A          & BPL & -0.66$_{-0.93}^{+0.43}$       & -3.07$_{-0.64}^{+0.53}$       & --                          & -0.60$_{-0.40}^{+0.50}$      & 0.0015$_{-0.0001}^{+0.0004}$             & --                 \\
170728B          & BPL & -0.63$_{-0.06}^{+0.06}$       & -1.32$\pm 0.03$               & --                          & -0.15$_{-0.28}^{+0.32}$      & 0.034$\pm0.006$                          & --                 \\
180418A          & SPL & -0.98$_{-0.16}^{+0.12}$       & --                            & --                          & -0.85$\pm0.14$               & --                                       & --                 \\
180727A          & BPL & -1.95$_{-1.92}^{+0.25}$       & -0.67$_{-0.09}^{+0.22}$       & --                          & -0.61$_{-0.35}^{+0.36}$      & 0.019$_{-0.015}^{+1.016}$                & --                 \\
191031D          & SPL & -2.55$_{-0.27}^{+0.25}$       & --                            & --                          & -0.93$_{-0.28}^{+0.17}$      & --                                       & --                 \\
200219A          & BPL & -5.66$_{-0.22}^{+0.25}$       & -1.33$_{-0.18}^{+0.14}$       & --                          & -0.83$_{-0.19}^{+0.39}$      & 0.0062$_{-0.0003}^{+0.0129}$             & --                 \\
201006A          & SPL & -1.01$_{-0.08}^{+0.07}$       & --                            & --                          & -1.10$\pm0.55$               & --                                       & --                 \\
210726A          & SPL & -0.72$\pm0.02$                & --                            & --                          & -0.71$\pm 0.11$              & --                                       & --                 \\
\enddata
\tablecomments{Column description: (1) GRB name. (2) Best-fit model for the X-ray afterglow. The abbreviations ``SPL'', ``BPL'' and ``TPL'' represent the single, broken, or triple power-law models. ($3-5$) Temporal decay indices for each segment of the best-fit model. (6) X-ray spectral index. ($7-8$) Time (in days) of the breaks for the BPL and TPL models. In case of SGRBs with detected jet breaks in their light curves, we identify the time of the jet break (Section\,\ref{sec:opening_angles}) with $t_{b1}$ for the broken power-law model and $t_{b2}$ for the triple power-law model. Errors are 1$\sigma$.\\
$^{*}$ Value derived from \textit{Swift} photon index ($\Gamma_{X}$). \\
$^{\dagger}$ The radio afterglow light curve of GRB~211106A shows a jet break at $t_{\rm j} \approx 29$ days \citep{Laskar2022}.\\
}
\label{Table:LC_fitting_parameters}
\end{deluxetable*}
%

%
\begin{longrotatetable}
\begin{deluxetable*}{cccccccccccc}
\tabletypesize{\footnotesize}
\tablecaption{Inferred Properties, Jet Opening Angles Measurements and Lower Limits of the SGRB Sample}
\setlength{\tabcolsep}{3pt}
\linespread{1.3}
\tablehead{
\colhead{GRB} & \colhead{$\nu_c<\nu_x$} & \colhead{p} & \colhead{$\epsilon{_B}$}& \colhead{$E_{\gamma\rm{,iso}}$} &\colhead{$\langle E_{\rm{K,iso}} \rangle$} & \colhead{$\langle n_0 \rangle$} & \colhead{$\langle \theta_{\rm j} \rangle$ or $\langle \theta_{\rm j, min} \rangle$} & \colhead{$f_b$}& \colhead{$E_{\gamma}$} &\colhead{E$_{\rm{K}}$} & \colhead{$\eta_{\gamma}$} \\
 \colhead{} & \colhead{} & \colhead{} & \colhead{}& \colhead{(erg)} &\colhead{(erg)} & \colhead{(cm$^{-3}$)} & \colhead{($^{\circ}$)} & \colhead{}& \colhead{(erg)} &\colhead{(erg)} & \colhead{}
}
\startdata
\multicolumn{12}{c}{\textbf{Jet Opening Angle Detections}} \\
\hline
050724A & N & 2.29$\pm0.10$   & 10$^{-4}$ & (1.2$\pm 0.2$)$\times10^{51}$         & (1.8$_{-0.5}^{+0.8}$)$\times10^{51}$       & (8.9$_{-4.9}^{+5.9})\times10^{-1}$       & 33.9$_{-4.6}^{+3.5}$     & 1.7$\times10^{-1}$   & 2.0$\times10^{50}$ & 3.1$\times10^{50}$ & 0.39 \\
051221A & Y & 2.24$\pm0.07$   & 0.1       & (8.6$_{-0.4}^{+0.5}$)$\times10^{51}$  & (2.7$\pm 0.3$)$\times10^{51}$              & (1.4$_{-0.4}^{+0.5})\times10^{-1}$       & 6.9$_{-0.7}^{+0.8}$      & 7.2$\times10^{-3}$   & 6.2$\times10^{49}$ & 2.0$\times10^{49}$ & 0.76 \\ 
111020A & Y & 2.08$\pm0.32$   & 0.1       & (3.9$_{-1.0}^{+1.2}$)$\times10^{50}$  & (4.8$_{-0.8}^{+0.9}$)$\times10^{51}$       & (4.6$_{-0.8}^{+1.4})\times10^{-3}$       & 8.0$_{-1.4}^{+1.7}$      & 1.0$\times10^{-2}$   & 3.8$\times10^{48}$ & 4.7$\times10^{49}$ & 0.08 \\
130603B & Y & 2.70$\pm0.06$   & 0.1       & (2.9$\pm 0.2$)$\times10^{51}$         & (1.1$_{-0.1}^{+0.2}$)$\times10^{51}$       & 0.09$_{-0.03}^{+0.04}$                   & 3.7$\pm0.3$              & 2.1$\times10^{-3}$   & 6.0$\times10^{48}$ & 2.3$\times10^{48}$ & 0.73 \\   
140903A & N & 2.27$\pm0.16$   & 10$^{-3}$ & (2.9$\pm 0.3$)$\times10^{50}$         & (2.9$_{-0.7}^{+0.9}$)$\times10^{52}$       & (0.3$_{-0.2}^{+0.3})\times10^{-2}$       & 3.2$_{-2.0}^{+0.8}$      & 1.6$\times10^{-3}$   & 4.6$\times10^{47}$ & 4.5$\times10^{49}$ & 0.01 \\
150424A & N & 2.40$\pm0.15$   & 0.01      & (3.7$\pm 0.6$)$\times10^{51}$         & (3.4$_{-1.8}^{+4.0})\times10^{51}$         & (6.4$_{-5.3}^{+28.8})\times10^{-4}$      & 4.3$_{-2.5}^{+2.1}$      & 2.8$\times10^{-3}$   & 1.0$\times10^{49}$ & 9.6$\times10^{48}$ & 0.52 \\ 
160821B & Y & 2.36$\pm0.14$   & 0.01      & (3.3$_{-0.4}^{+0.5}$)$\times10^{49}$  & (1.5$_{-0.7}^{+1.8})\times10^{50}$         & (1.7$_{-1.4}^{+6.5})\times10^{-2}$       & 8.4$_{-3.2}^{+4.9}$      & 1.1$\times10^{-2}$   & 3.5$\times10^{47}$ & 1.6$\times10^{48}$ & 0.18 \\
200411A & N & 2.10$\pm0.07$   & 0.1       & (2.0$_{-0.4}^{+0.5}$)$\times10^{51}$  & (2.4$_{-1.1}^{+2.2})\times10^{52}$         & (1.7$_{-1.3}^{+6.2})\times10^{-5}$       & 1.6$_{-0.5}^{+0.7}$      & 3.9$\times10^{-4}$   & 7.8$\times10^{47}$ & 9.4$\times10^{48}$ & 0.08 \\ 
200522A & N & 1.78$\pm0.30$   & 0.01      & (4.5$_{-1.0}^{+1.9}$)$\times10^{50}$  & (4.5$_{-2.6}^{+6.0})\times10^{51}$         & (3.4$_{-2.7}^{+11.9})\times10^{-3}$      & 6.2$_{-1.8}^{+2.4}$      & 5.8$\times10^{-3}$   & 2.7$\times10^{48}$ & 2.6$\times10^{49}$ & 0.09 \\
$^{\dag}$211106A & N & 2.47$\pm0.05$   & $~10^{-5}$ & (4.4$\pm0.4$)$\times10^{51}$ & (1.6$_{-0.9}^{+1.7})\times10^{53}$         & 0.3$\pm 0.1$       & 15.5$\pm1.4$     & 3.9$\times10^{-2}$   & 1.7$\times10^{50}$ & 6.3$\times10^{51}$ & 0.03 \\
\hline
\hline
\multicolumn{12}{c}{\textbf{Jet Opening Angle Non-Detections}} \\
\hline
050509B          & N & 2.14$\pm0.83$   & 0.1       & (9.1$_{-3.6}^{+5.0}$)$\times10^{48}$ & (1.2$_{-0.9}^{+3.0})\times10^{50}$         & (7.8$_{-7.4}^{+142.7})\times10^{-5}$      & 2.6$_{-1.1}^{+1.8}$     & 1.0$\times10^{-3}$   & 9.3$\times10^{45}$ & 1.2$\times10^{47}$ & 0.07 \\   
$^{*}$050709     & Y & 2.31$\pm0.13$   & 0.1       & (2.8$_{-0.2}^{+0.4}$)$\times10^{49}$ & (2.9$_{-0.5}^{+0.8})\times10^{49}$         & (8.4$_{-4.0}^{+5.2})\times10^{-1}$        & 25.8$_{-2.6}^{+2.2}$    & 1.0$\times10^{-1}$   & 2.8$\times10^{48}$ & 2.9$\times10^{48}$ & 0.49 \\   
100117A          & Y & 2.36$\pm0.30$   & 0.1       & (2.9$_{-0.7}^{+0.8}$)$\times10^{51}$ & (1.9$_{-0.3}^{+0.3})\times10^{51}$         & (0.4$_{-0.1}^{+0.3})\times10^{-1}$        & 10.3$_{-0.6}^{+0.9}$    & 1.6$\times10^{-2}$   & 4.6$\times10^{49}$ & 3.1$\times10^{49}$ & 0.60 \\   
101219A          & N & 2.73$\pm0.13$   & 0.1       & (9.9$_{-1.1}^{+1.2}$)$\times10^{51}$ & (0.5$_{-0.4}^{+1.2})\times10^{52}$         & (0.2$_{-0.2}^{+4.1})\times10^{-4}$        & 0.5$_{-0.2}^{+0.4}$     & 3.8$\times10^{-5}$   & 3.8$\times10^{47}$ & 1.9$\times10^{47}$ & 0.66 \\ 
110112A          & Y & 2.49$\pm0.07$   & 0.1       & (1.3$_{-0.5}^{+0.6}$)$\times10^{50}$ & (6.5$_{-0.6}^{+0.6})\times10^{50}$         & (2.3$\pm0.4)\times10^{-2}$                & 4.8$\pm0.2$             & 3.5$\times10^{-3}$   & 4.4$\times10^{47}$ & 2.3$\times10^{48}$ & 0.16 \\
111117A          & Y & 2.27$\pm0.07$   & 0.1       & (3.1$_{-0.7}^{+0.9}$)$\times10^{52}$ & (1.7$_{-0.3}^{+0.4})\times10^{51}$         & (1.3$_{-0.4}^{+0.9})\times10^{-1}$        & 9.2$_{-0.6}^{+0.8}$     & 1.3$\times10^{-2}$   & 4.0$\times10^{50}$ & 2.2$\times10^{49}$ & 0.95 \\   
120804A          & Y & 2.08$\pm0.11$   & 0.1       & (2.5$\pm 0.2$)$\times10^{52}$        & (1.1$_{-0.2}^{+0.3})\times10^{52}$         & (0.3$_{-0.1}^{+0.3})\times10^{-2}$         & 10.5$_{-1.0}^{+1.2}$    & 1.7$\times10^{-2}$   & 4.2$\times10^{50}$ & 1.8$\times10^{50}$ & 0.69 \\
140930B          & N & 2.67$\pm0.19$   & 0.1       & (3.9$_{-0.8}^{+0.9}$)$\times10^{52}$ & (0.4$_{-0.3}^{+1.0})\times10^{52}$         & (0.2$_{-0.2}^{+4.5})\times10^{-4}$        & 4.1$_{-1.8}^{+3.1}$     & 2.6$\times10^{-3}$   & 1.0$\times10^{50}$ & 1.0$\times10^{49}$ & 0.91 \\   
150101B          & N & 2.22$\pm0.01$   & 0.1       & (3.0$_{-2.2}^{+2.3}$)$\times10^{47}$ & (4.9$_{-2.3}^{+4.2})\times10^{51}$         & (1.0$_{-0.8}^{+3.4})\times10^{-5}$        & 9.4$_{-2.2}^{+2.8}$     & 1.3$\times10^{-2}$   & 4.0$\times10^{45}$ & 6.6$\times10^{49}$ & $6\times10^{-5}$ \\   
150423A          & N & 2.32$\pm0.01$   & 0.1       & (5.0$_{-1.3}^{+1.6}$)$\times10^{51}$ & (7.4$_{-4.2}^{+9.6})\times10^{51}$         & (3.0$_{-2.6}^{+18.9})\times10^{-5}$      & 0.7$_{-0.2}^{+0.3}$     & 7.5$\times10^{-5}$   & 3.7$\times10^{47}$ & 5.5$\times10^{47}$ & 0.4 \\ 
150831A          & N & 2.43$\pm0.09$   & 0.1       & (2.1$_{-0.3}^{+0.4}$)$\times10^{52}$ & (7.4$_{-3.8}^{+9.2})\times10^{51}$         & (4.1$_{-3.6}^{+20.2})\times10^{-5}$       & 0.8$_{-0.2}^{+0.3}$     & 1.0$\times10^{-4}$   & 2.0$\times10^{48}$ & 7.2$\times10^{47}$ & 0.74 \\    
160624A          & N & 3.81$\pm0.60$   & 0.1       & (5.4$_{-2.1}^{+3.6}$)$\times10^{50}$ & (3.7$_{-2.1}^{+4.8})\times10^{51}$         & (5.1$_{-4.7}^{+56.6})\times10^{-5}$       & 0.3$_{-0.1}^{+0.2}$     & 1.4$\times10^{-5}$   & 7.5$\times10^{45}$ & 5.1$\times10^{46}$ & 0.13 \\  
170728B          & N & 2.75$\pm0.03$   & 0.1       & (3.1$_{-0.8}^{+0.9}$)$\times10^{52}$ & (1.5$_{-0.7}^{+1.2})\times10^{53}$         & (3.3$_{-2.6}^{+11.8})\times10^{-5}$       & 3.5$_{-0.8}^{+1.1}$     & 1.9$\times10^{-3}$   & 5.8$\times10^{49}$ & 2.8$\times10^{50}$ & 0.17 \\ 
180418A          & N & 2.39$\pm0.12$   & 0.1       & (2.7$_{-1.4}^{+1.5}$)$\times10^{51}$ & (3.7$_{-2.2}^{+4.6})\times10^{51}$         & (2.6$_{-0.9}^{+2.2})\times10^{-4}$        & 11.5$_{-1.3}^{+2.1}$    & 2.0$\times10^{-2}$   & 5.4$\times10^{49}$ & 7.4$\times10^{49}$ & 0.42 \\ 
180727A          & N & 1.92$\pm0.20$   & 0.1       & (1.0$_{-0.1}^{+0.2}$)$\times10^{51}$ & (2.3$_{-1.0}^{+1.8})\times10^{52}$         & (1.8$_{-1.4}^{+6.5})\times10^{-5}$        & 3.0$_{-0.7}^{+0.9}$     & 1.4$\times10^{-3}$   & 1.4$\times10^{48}$ & 3.2$\times10^{49}$ & 0.04 \\ 
191031D          & N & 2.88$\pm0.45$   & 0.01      & (5.9$_{-1.0}^{+1.2}$)$\times10^{51}$ & (2.3$_{-1.8}^{+7.7})\times10^{51}$         & (3.6$_{-3.5}^{+188.9})\times10^{-4}$      & 0.6$_{-0.3}^{+0.6}$     & 5.5$\times10^{-5}$   & 3.2$\times10^{47}$ & 1.2$\times10^{47}$ & 0.72 \\  
200219A          & N & 2.76$\pm0.21$   & 0.01      & (7.6$_{-1.9}^{+2.3}$)$\times10^{51}$ & (2.8$_{-2.1}^{+9.3})\times10^{51}$         & (4.1$_{-3.9}^{+173.1})\times10^{-4}$      & 2.0$_{-1.0}^{+1.8}$     & 6.1$\times10^{-4}$   & 4.6$\times10^{48}$ & 1.7$\times10^{48}$ & 0.73 \\
201006A          & Y & 2.01$\pm0.10$   & 0.1       & (4.3$_{-0.7}^{+0.6}$)$\times10^{50}$ & (1.6$_{-0.3}^{+0.4})\times10^{51}$         & 0.8$_{-0.7}^{+6.4}$                       & 3.4$_{-0.8}^{+1.1}$     & 1.8$\times10^{-3}$   & 7.5$\times10^{47}$ & 2.8$\times10^{48}$ & 0.21 \\ 
210726A          & N & 1.97$\pm0.03$   & 0.1       & (7.7$\pm 1.9$)$\times10^{49}$        & (1.4$_{-0.6}^{+1.0})\times10^{52}$         & (0.7$_{-0.4}^{+1.1})\times10^{-5}$        & 5.1$_{-0.9}^{+1.1}$     & 4.0$\times10^{-3}$   & 3.0$\times10^{47}$ & 5.5$\times10^{49}$ & 0.01 
\enddata
\tablecomments{Column description: (1) GRB name. (2) ``Y'' (yes) and ``N'' (no) indicate whether or not $\nu_{\rm c}$ is above $\nu_{\rm X}$. (3) Power-law index of the input distribution of accelerated electron. (4) Post-shock energy fractions transmitted to the magnetic field ($\epsilon_{\rm B}$). (5) The isotropic $\gamma$-ray energy ($\langle E_{\gamma, {\rm iso}} \rangle$) values. (6) The median values of the isotropic-kinetic energy ($\langle E_{\rm{K,iso}} \rangle$). (7) The median values of the circumburst denstity ($\langle n \rangle$). (8) Median values of the jet opening angles detections and lower limits represented by $\theta_{\rm j}$ and $\theta_{\rm j, min}$. (9) The beaming correction factor ($f_{\rm b}$). (10) True beaming corrected $\gamma$-ray energy (erg). (11) True beaming corrected kinetic energy (erg). (12) The $\gamma$-ray efficiency ($\eta_{\gamma}$) calculated as $\eta_{\gamma}= E_{\gamma, {\rm iso}} / (E_{\gamma, {\rm iso}}+E_{\rm k, iso})$.\\
We use $E_{\gamma, {\rm iso}}= k_{bol} \times 4 \pi f_{\gamma}d_L^2(1+z)^{-1}$ where $k_{bol}$ is the bolometric correction factor to convert the \textit{Swift}/BAT fluence ($f_{\gamma}$) in the $15-150$\,keV energy band to the $1-10000$\,keV energy band ($k_{bol}=5$), and $d_L$ is the luminosity distance (for SGRBs with unknown redshifts in Table\,\ref{Table:log_GRB_BasicInformation} we assign the fiducial value of $z=0.5$). The isotropic-kinetic energy and circumburst density are calculated with the fraction of the post-shock energy transmitted to electrons ($\epsilon_{\rm e}$) fixed to 0.1 for the different values of $\epsilon_{\rm B}$ and redshifts (Table\,\ref{Table:log_GRB_BasicInformation}). We define $f_{\rm b}$ as $f_{b}=1 - \cos(\theta_{\rm j})$. Uncertainties are 1$\sigma$.\\
$^{*}$The $E_{\gamma\rm{,iso}}$ value of GRBs~050709 is given in the $1-10000$\,keV energy band by \citet{Villasenor2005}. $^{\dag}$ Values obtained from the analysis of \citet{Laskar2022} considering $z=1$ and IC/KN corrections.
}
\label{Table:energy_density}
\end{deluxetable*}
\end{longrotatetable}
%

\subsection{Isotropic-equivalent Kinetic Energies and Circumburst Densities}  \label{sec:energy_density_analysis}
Here we constrain the values of $p$, $E_{\rm K,iso}$ and $n_0$ as described in Section~\ref{sec:comp_broadband_analysis} using any available X-ray, optical, NIR and radio data in the framework of the standard synchrotron forward shock model \citep{Sari1998,Granot2002}; these are key parameters which feed into the determination of the opening angles. For the SGRBs detected during 2005-2015, we utilize the broad-band analysis from \citet{Fong2015a} for the values of $p$, $E_{\rm K,iso}$ and $n_0$. We also use the analysis of \citet{Schroeder2020} to determine these parameters for the particular cases of GRBs~050509B, 150424A and 160821B, and the latest analysis of GRB~211106A from \citet{Laskar2022}. For the remaining SGRBs discovered beyond 2015, we follow the methods of \citet{Fong2015a} to constrain these shock and environment parameters.

First, we constrain the position of $\nu_{\rm X}$ with respect to $\nu_c$ (e.g., $\nu_{\rm X} > \nu_{\rm c}$ or $\nu_{\rm m} < \nu_{\rm X} < \nu_{\rm c}$), assuming a constant density medium. We determine under which scenario the values of $p$, independently determined from $\alpha_X$ and $\beta_X$, adhere to the closure relations given by \citet{Granot2002}. For each SGRB, we require that the values of $p$ derived from $\alpha_X$ and $\beta_X$ to be consistent within $1\sigma$, calculate the weighted mean (Table~\ref{Table:energy_density}) and assume this value as $p$. For two bursts, GRBs~180727A and 210726A, we obtain $\langle p \rangle =1.92\pm0.20$ and $\langle p \rangle =1.97\pm0.03$, respectively, which yield divergent total integrated electron energies. Thus we use $p$ derived from $\beta_X$ for these cases as this parameter is generally more reliable since it remains constant over time, which both give $p>2$. Additionally, in the case of GRB~200522A we find a weighted mean of $\langle p \rangle =1.78\pm0.30$; therefore we adopt $p = 2.05 \pm 0.03$ reported in \citet{Fong2021} for our analysis. Deriving from the standard closure
relations, we find $\nu_{\rm X} > \nu_{\rm c}$ for 10 cases and $\nu_{\rm X} < \nu_{\rm c}$ for 19 cases, with an inferred weighted mean of $\langle p \rangle = 2.30 \pm 0.32$ for the total sample of 29 SGRBs. This value is in agreement with the $\langle p \rangle = 2.40_{-0.36}^{+0.28}$ found by \citet{Fong2015a}.

Next, we use the broad-band data to calculate the physical burst properties, $E_{\rm K,iso}$ and $n_0$, following the relations of the synchrotron model \citep{Granot2002}, which provides a mapping from the light curve and SEDs to physical properties. We assume that the radio band follows $\nu_{\rm sa}<\nu_{\rm rad}<\nu_{m}$, and that the optical band is located at $\nu_{\rm m} < \nu_{\rm opt}<\nu_{\rm c}$. We utilize the location of $\nu_c$ as an additional constraint. If $\nu_{\rm c}<\nu_{\rm X}$, we set the maximum value to $\nu_{\rm c, max}=7.3 \times 10^{16}$\,Hz, at the lower edge of the X-ray band (0.3~keV). If $\nu_{m}<\nu_{\rm X}<\nu_{\rm c}$, we constrain the minimum value of $\nu_{\rm c,min}=2.4 \times 10^{18}$\,Hz corresponding to the upper edge of the X-ray band (10~keV).

In general, we first fix $\epsilon_{\rm e}$ and $\epsilon_{\rm B}$ to the equipartition values of $\epsilon_{\rm e}=0.1$ and $\epsilon_{\rm B}=0.1$\footnote{We note that the values of the isotropic-equivalent kinetic energies and circumburst densities are degenerate and depend on the assumptions on the microphysical parameters we choose. However, since we are taking into account the entire distributions of $E_{\rm K,iso}$ and $n_0$, we find that the values of $\theta_{\rm j}$ are consistent within the errors independent of the $\epsilon_{\rm B}$ value we assume.} (Table\,\ref{Table:energy_density}). This choice of microphysical parameters is based on results from GRB afterglow fitting \citep{Ryan2015,Beniamini2017} and theoretical studies which show that part (typically $\sim10\%$) of the kinetic energy may be deposited into non-thermal particles \citep{Spitkovsky2008}. For eight cases, we need lower values of $\epsilon_{\rm B}=10^{-4}-0.01$ in order to find $E_{\rm K,iso}$ and $n_0$ solution pairs \citep[e.g.,][]{Panaitescu2002,Sironi2011}. In particular, these cases are 150424A, 160821B \citep{Schroeder2020}, 191031D, 200219A and 200522A \citep{Fong2021} with $\epsilon_{\rm B}=0.01$, GRB~140903A with $\epsilon_{\rm B}= 10^{-3}$ \citep{Fong2015a}, GRBs~050724A with $\epsilon_{\rm B}= 10^{-4}$ \citep{Fong2015a} and 211106A with $\epsilon_{\rm B}= 10^{-5}$ \citep[assuming $z=1$ and IC/KN corrections][]{Laskar2022}. Additionally, we set $z=0.5$ for SGRBs with no redshift information (Table\,\ref{Table:log_GRB_BasicInformation}). We point out that the derived physical parameters will be degenerate with respect to the fraction of particles accelerated into a non-thermal distribution \citep{Eichler2005}.

To calculate the distributions of $E_{\rm K,iso}$ and $n_0$, we consider a parameter space with 1000 logarithmically-spaced steps for each parameter, with ranges of $E_{\rm K,iso}=10^{46}-10^{56}$\,erg and $n_0=10^{-6}-10^{3}$\,cm$^{-3}$, where $n_{\rm min}=10^{-6}$\,cm$^{-3}$ corresponds to the typical value of the intergalactic medium. We then constrain the $E_{\rm K,iso}-n$ parameter space by combining the probability distributions from each wavelength for each SGRB, marginalize in each parameter, and normalize the 1D distributions. In Table\,\ref{Table:energy_density} we report on the values of $E_{\rm K,iso}$ and $n_0$ ($1\sigma$ uncertainties) that vary between [$3\times10^{49}-1.6\times10^{53}$]\,erg and [$0.7\times10^{-5}-0.8$]\,cm$^{-3}$, respectively. The median values for the final sample of 29 SGRBs are $\langle E_{\rm{K, iso}}\rangle \approx 3.7 \times 10^{51}$\,erg and $\langle n_0 \rangle \approx 6.4\times 10^{-4}$\,cm$^{-3}$ considering the equipartition parameters in Table\,\ref{Table:energy_density}.

\section{Jet opening Angles, Beaming-Corrected Energetics and Event Rates} \label{sec:opening_angles}
As a consequence of the relativistic beaming, the observed temporal behavior from a spherical expansion for an on-axis observer is initially similar to that of a highly collimated relativistic outflow. As the jet interacts with the circumburst medium, the value of its bulk Lorentz factor ($\Gamma$) declines over time to reach a value equal to the inverse of the jet opening angle \citep[$\Gamma\sim\theta_{\rm j}^{-1}$;][]{Piran1995} and consequently the edge of the highly collimated relativistic outflow becomes visible to the observer. This moment is known as the time of the jet break ($t_{\rm j}$) and observationally manifests as an achromatic and temporal steepening in the afterglow light curve \citep{Sari1999,vanEerten2013}, after which the outflow may follow lateral expansion \citep{Granot2012}. At this point, the jet undergoes a lateral expansion and the afterglow decline rate can be analytically described by $t^{-p}$ \citep{Rhoads1999,Sari1999}. By constraining the parameter $t_{\rm j}$ in the SGRB afterglow light curves, one can additionally derive $\theta_{\rm j}$.

\subsection{Determination of Jet Opening Angles}
To identify jet breaks from the best-fit models, we define a jet break as the time when $\alpha$ significantly steepens. Quantitatively, for $\alpha_{\rm X, i} > \alpha_{\rm X, i+1}$, we classify light curves as having jet breaks if they exhibit: (i) $\alpha_{\rm X, i+1}<-1.75$, and (ii) $0.75 < |\Delta \alpha_{\rm X}|< 3$. For instance, considering an ISM-like medium for the closure relations given by \citet{Granot2002} and assuming $2<p<3$ \citep{Fong2015a}, we know that $\alpha_{\rm X}$ can range between $-1.5<\alpha_{\rm X}<-0.75$ if $\nu_{\rm X}<\nu_{\rm c}$ and $-1.75<\alpha_{\rm X}<-1$ if $\nu_{\rm c}<\nu_{\rm X}$, in case of a spherical outflow. Therefore, if the post-break segment of a light curve is steeper than $\alpha_{\rm X}=-1.75$, the only real justification is a jet break instead of the transition of $\nu_{\rm c}$ across the X-ray band. Additionally, the transition of $\nu_{\rm c}$ across the X-ray band predicts a maximum value of the change in slope of $|\Delta \alpha_{\rm X}|=0.25$ \citep{Sari1998}, so that any larger $\Delta \alpha$ is most naturally explained by a jet break.

\begin{figure*}
\centering
\includegraphics[width=0.8\textwidth]{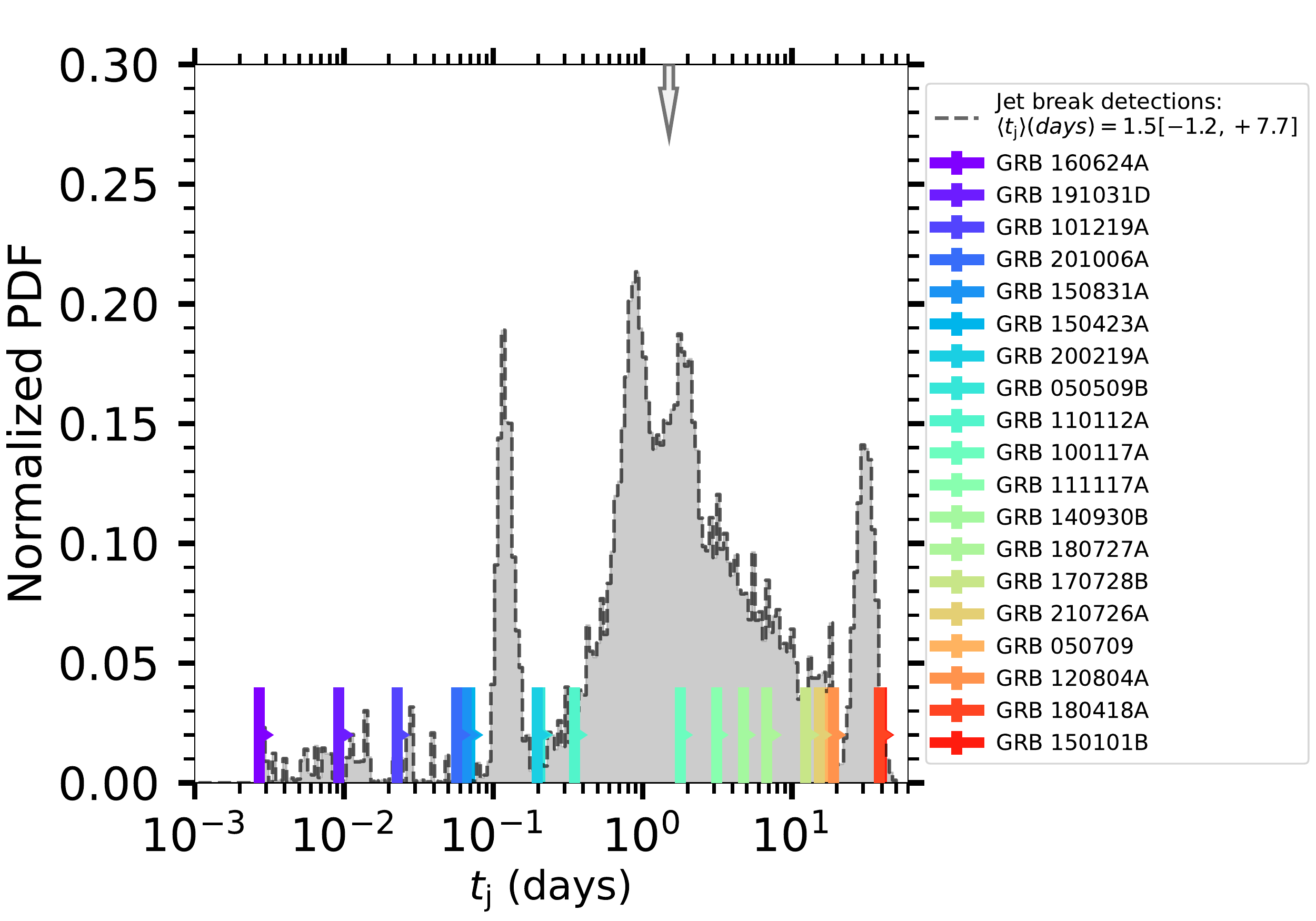}
\vspace{-0.1in}
\caption{Distribution of the jet break times for 10 SGRBs with jet break detections (grey histogram) derived from the individual jet break time posteriors. The grey arrow indicates the median jet break time of $\langle t_{\rm j} \rangle = 1.5[-1.2,+7.7]$~days. The distribution at $t_{\rm j}>20$\,days is clearly dominated by a peak which represents the late-time detection of the GRB~211106A jet break in the radio band. For the 19 lower limits, the times of the last X-ray detections are indicated with color-coded lower limits which roughly corresponds to the time out to which we are sensitive to jet breaks for each event. A significant fraction  (10/19) of the lower limits are found at times of $t_{\rm j}>\langle t_{\rm j} \rangle$, and specifically there are 6 events with last X-ray detections at $t_{\rm j}>10$\,days. This shows the existence of a SGRB group with jet breaks located at very late times (and potential wide opening angles) as the case of GRB~211106A.}
\label{fig:PDF_jet_break_times}
\end{figure*}

Considering an ISM-like medium and a top-hat geometry for the cosmological SGRB jets with the observer along the jet axis\footnote{One of the lessons learnt from GRB\,170817A is that jets can be more complex and structured. \cite{Wu2019} showed that cosmological SGRBs share a similar jet structure to GRB\,170817A. Likewise, \cite{Urrutia2021} demonstrated that structured jets can have significantly more injected energy at large observing angles than top-hat jets. However, for our SGRB sample, the emission is completely dominated by the jet core making the detection of any structure difficult. Therefore, the assumption of a SGRB top-hat jet geometry is a reasonable approximation for our study.}, we combine the information from the X-ray afterglow light curves with constraints on E$_{\rm K,iso}$ and $n_0$ (Section~\ref{sec:comp_broadband_analysis}) to calculate $\theta_{\rm j}$ or their lower limits. We use the following equation given by \citet{Sari1999,Frail2001}:

\begin{equation}
    \theta_{\rm j} = 9.5 t_{\rm j}^{3/8} (1+z)^{-3/8} E_{\rm K,iso,52}^{-1/8} n{_0}^{1/8}~~~{\rm [deg]}
\label{eqn:jb}
\end{equation}

\noindent where $t_{\rm j}$ is time in days, E$_{\rm{K, iso, 52}}$ is in units of $10^{52}$\,erg and $n_0$ is in units of cm$^{-3}$. For bursts with detected jet breaks we set $t_{\rm j}$ to the best-fit time of the break in our calculation of $\theta_{\rm j}$ (Table\,\ref{Table:LC_fitting_parameters}). In the case of SGRBs with no detected jet breaks (e.g., well-described by single power-law declines), we fix $t_{\rm j}$ to the time of the last X-ray detection (considering this time as a lower limit on $t_{\rm j}$) to obtain lower limits on the jet opening angles (Figure\,\ref{fig:PDF_jet_break_times}).

%
%

\subsection{Distribution of SGRB Jet Opening Measurements} \label{sec:distribution_opening_angles}
Using the aforementioned criteria, we find that nine SGRBs (SGRBs~050724A, 051221A, 111020A, 130603B, 140903A, 150424A, 160821B, 200411A and 200522A; Table\,\ref{Table:LC_fitting_parameters}) exhibit temporal steepenings in their X-ray afterglow light curves that we attribute to jet breaks. We also add to this group the case of GRB~211106A, for which the jet break was uncovered in the radio afterglow light curve \citep[][]{Laskar2022}. Thus, our final subsample is composed of 10 SGRBs with jet breaks. We confirme the existence of the jet breaks for eight cases that have been already published: GRB~051221A \citep{Burrows2006,Soderberg2006}, GRB~111020A \citep{Fong2012}, GRB~130603B \citep{Fong2014}, 140903A \citep{Troja2016}, GRB~150424A \citep{Jin2018}, GRB~160821B \citep{Lamb2019,Troja2019}, GRB~200522A \citep{OConnor2021} and GRB~211106A \citep{Laskar2022}. However, we update the time of the jet breaks on the X-ray afterglow light curves for all these cases based on our uniform modeling (see Table~\ref{Table:energy_density}), except the jet break time of GRB~211106A that was detected in the radio band. Additionally, we report on two new jet break detections in X-rays that have not been found to date: GRBs~050724A and 200411A (see Table~\ref{Table:energy_density}). For these ten SGRBs, we find that their median jet break times span a range of $t_{\rm j} \approx 0.1-30$~days. We use the posteriors of $t_{\rm j}$ for each case to build the total probability distribution of jet break times (Figure\,\ref{fig:PDF_jet_break_times}), and derive a median value of $\langle t_{\rm j} \rangle = 1.5[-1.2,+7.7]$~days.

\begin{figure*}[!t]
\centering
\includegraphics[width=0.49\textwidth]{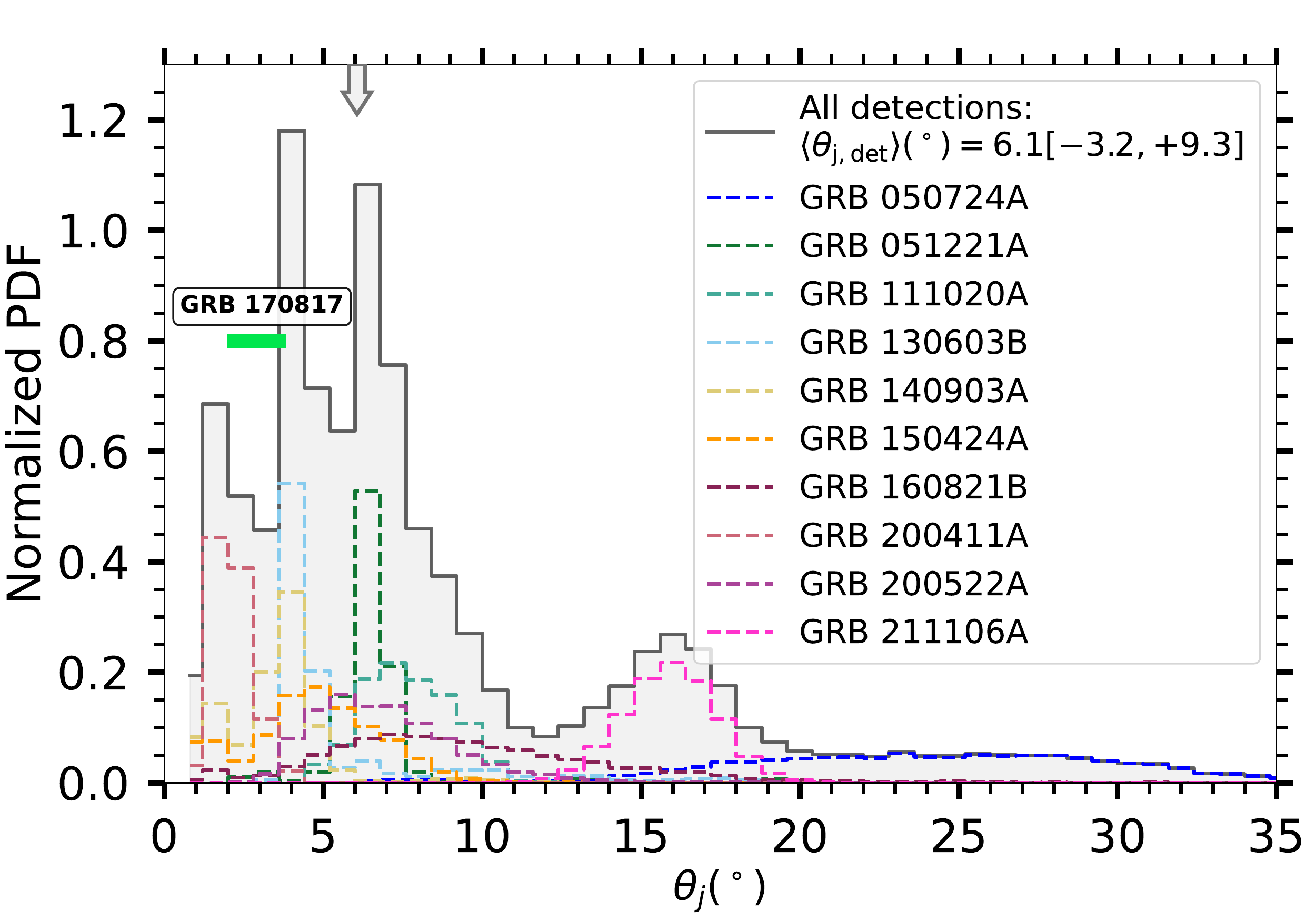}
\includegraphics[width=0.49\textwidth]{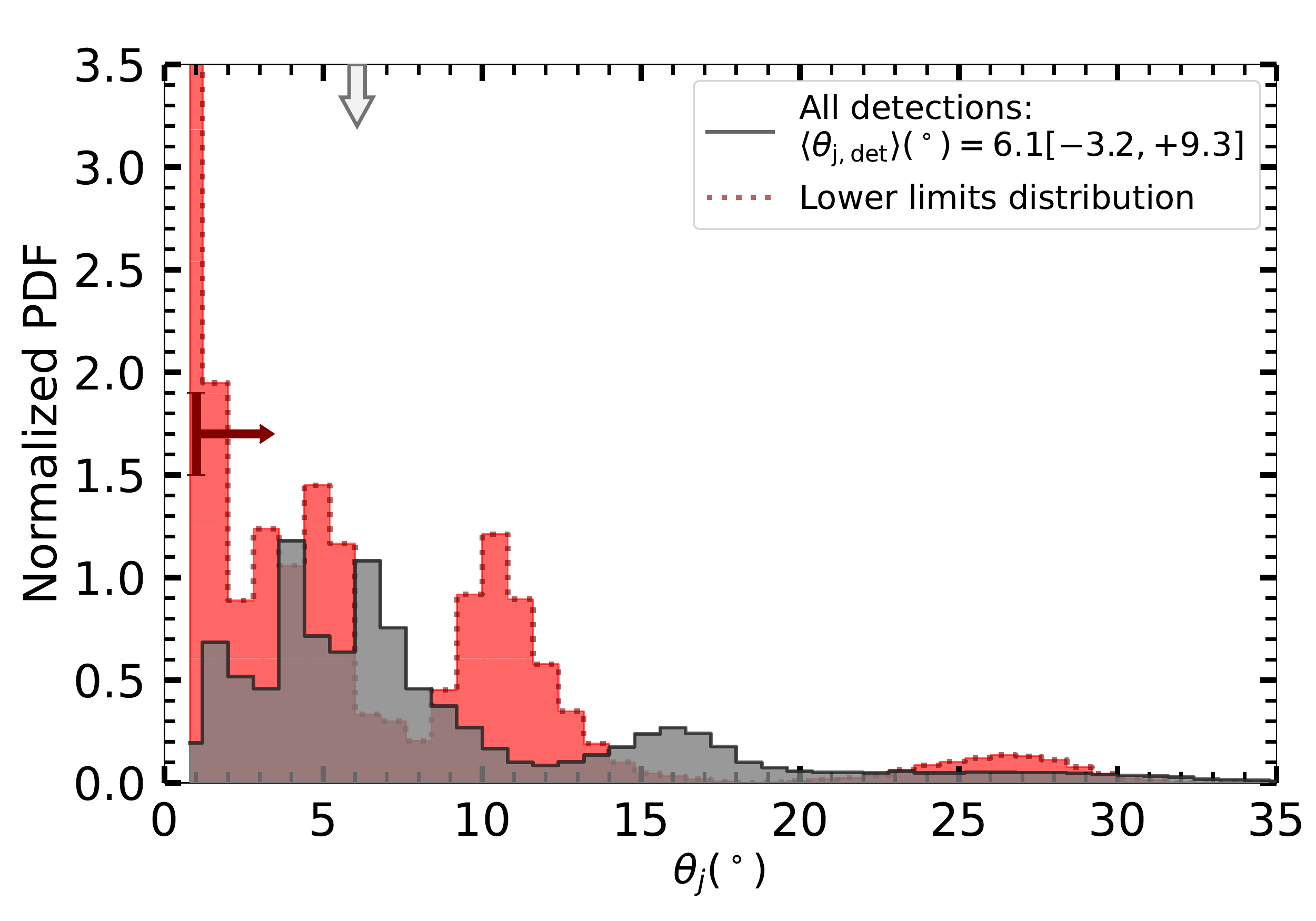}
\vspace{-0.1in}
\caption{\textit{left:} The normalized probability distribution for each of the 10 SGRBs with detected jet opening angles (color-coded open histograms), along with the total distribution of all 10 events (grey histogram). For this sample, we find a median value of $\langle \theta_{\rm j,det} \rangle=6.1^{\circ}[-3.2^{\circ},+9.3^{\circ}]$ (grey arrow). The inferred core jet opening angle for SGRB~170817A, $\theta_{\rm j,core} = 2^{\circ}-4^{\circ}$ \citep{Margutti2021}, is indicated as a green bar. \textit{right:} The total probability distribution for the sample of 19 lower limits (red dashed histogram) in comparison with the total distribution of detected opening angles (dark grey histogram). The red arrow indicates that this distribution represents lower limit values. There is a distinctive group of SGRB lower limits composed of 6 events with wider minimum opening angles of $\theta_{\rm j,min}\gtrsim 10^{\circ}$. Thus, this indicates a possible population of SGRBs with wider opening angles than described from the distribution of measurements alone.}
\label{fig:PDF_detected_opening_angles}
\end{figure*}

To build the total distributions of jet opening angle measurements, we perform 5000 random draws from the probability distributions in E$_{\rm{K, iso}}$, $n_0$ and $t_{\rm j}$ posterior for each jet break detection case. Since the distributions of E$_{\rm{K, iso}}$ and $n_0$ are correlated, we select the appropriate value of $n_0$ for every drawn value of E$_{\rm{K, iso}}$. We then use Equation\,\ref{eqn:jb} to obtain individual posterior distributions of opening angles for the ten cases with detected jet breaks (Figure\,\ref{fig:PDF_detected_opening_angles}, \textit{left}). The median and $1\sigma$ confidence intervals for each jet measurement are reported in Table\,\ref{Table:energy_density}.

We find that the angles for GRBs~051221A, 111020A, 140903A, 160821B and 200522A are in agreement within the errors with those reported by \citealt{Burrows2006,Soderberg2006,Fong2012,Troja2019,OConnor2021}, respectively. However, we find an opening angle of $\sim 3^{\circ}$ for GRB~140903A narrower than $\theta_{j} \approx 5^{\circ}$ \citep{Troja2016}, and a narrower opening angle measurement of $\approx 4^{\circ}$ for GRB~150424A instead of $\theta_{j} \sim 7^{\circ}$ \citep{Jin2018}. For GRB~050724A, we now measure an opening angle of $\theta_{\rm j} \approx 34^{\circ}$, which is consistent with the previous wide opening angle lower limit of $\theta_{\rm j} \gtrsim 25^{\circ}$ \citep{Grupe2009}. The detection of a new jet break here is driven by the last \textit{Chandra} detection at $\delta t \approx 21.6$~days for which we find a $\sim0.8$ times fainter flux (see Table\,\ref{Table:log_Xray_observations}) than that reported by \citealt{Grupe2006}. This measurement is a limiting case of jet opening angles since theoretical studies of post-merger outflows derive a maximum value of $\theta_{\rm j,max}\approx30^{\circ}$ \citep{Ruffert1999,Aloy2005,Rosswog2005,Rezzolla2011,Lazzati2021}. 

In the case of GRB~130603B, \citet{Fong2014} found a jet break at $\delta t \approx 0.47$~days in the optical and radio afterglow light curves, but not in the X-ray band. However, our new MCMC modeling shows a jet break at $\approx 0.11$~days post-trigger in the X-ray light curve. This earlier jet-break time is driven by the correction of the contribution of the X-ray contaminant (Figure\,\ref{fig:panel_GRB130603B} in Appendix\,\ref{sec:panel_GRB130603B}) to the afterglow light curve. Therefore, one does not need to invoke extra energy injection from a magnetar to explain the X-ray flux level of late-time observations. We constrain the opening angle measurement of GRB~130603B to $\theta_{\rm j} \approx 4^{\circ}$, which is in agreement with the lower end of the opening angle range ($\approx 4^{\circ}-8^{\circ}$) for this event reported by \citet{Fong2014}.
 
For the jet opening angle measurements. Overall, we find a median value of $\langle \theta_{\rm j,det} \rangle=6.1^{\circ}[-3.2^{\circ},+9.3^{\circ}]$ for bursts with jet opening angle measurements, which is consistent with previous studies \citep{Fong2015a,Jin2018} and in line with the opening angle estimates of jets for SGRBs produced by BNS mergers \citep{Ghirlanda2016}. In the case of SGRB~170817A, the first bona-fide detection of an SGRB jet launched by a BNS merger, the range of values inferred for the core is $\theta_{\rm j,core} = 2^{\circ}-4^{\circ}$ \citep{Margutti2021}, which is narrower than the median value we find for the SGRB jet opening angle measurements. It is clear from Figure\,\ref{fig:PDF_detected_opening_angles} (\textit{left}) that the probability beyond $\theta_{\rm j} \gtrsim 15^{\circ}$ is dominated by two events, GRBs~050724A and 211106A.

%
%

\subsection{Distribution of Jet Lower Limits: A Population of Wider Jets} 
We also build probability distributions for the 19 SGRBs with inferred lower limits on the opening angles. We again perform 5000 random draws from the correlated distributions of E$_{\rm{K, iso}}$ and $n_0$. For these cases, we instead fix the jet break time to the (log-centered) time of the last X-ray detection. Using Equation\,\ref{eqn:jb}, we obtain the posterior distribution of the opening angle lower limit for each event, and report the median value of the minimum opening angle and $1\sigma$ confidence intervals (Table~\ref{Table:energy_density}). We find that the range of minimum opening angles varies between $\theta_{\rm j,min} \approx 0.3^{\circ}$-26$^{\circ}$.

We find that the minimum opening angles of 6 events (SGRBs~050709, 100117A, 111117A, 120804A, 150101B, and 180418A) are $\gtrsim 1\sigma$ larger than $\langle \theta_{\rm j,det} \rangle$, indicative of a population of wider jets. To investigate if the distribution of lower limits is statistically distinct from the distribution of jet measurements, we compare the CDFs at their 68\% credible regions. If the median of each distribution lies within the other's 68\% credible region, then the incorporation of the lower limit distribution does not add new information. Indeed, we find that the median values are within the 16th and 84th percentiles of each other's distribution. Thus, at first glance the distribution of jet measurements appear to be a fair approximation for the observed opening angles of SGRBs.

However, we acknowledge that this may be the consequence of bias due to the limited number of events observed at $t_{\rm j}\gtrsim 2-3$ days, for which we expect to have wider jet opening angles. Indeed, we have already detected two events with wide opening angle measurements and found six cases with minimum opening angles wider than $\theta_{\rm j,min} \gtrsim 10^{\circ}$. If more SGRBs were able to be monitored at late times (which would require more sensitive X-ray facilities than currently available), this could result in measurements or lower limits of broader jet opening angles. Therefore, for our subsequent analysis, we build a mock sample including a fiducial population of wider jets. In particular, we combine the events with opening angle measurements and six wide jets (representing the number of events with wider opening angle lower limits) with measurements centered around $\theta_{\rm j}\sim20^{\circ}$ \citep[motivated by the findings in][]{Murguia-Berthier2017} to investigate how an SGRB sample with constraining wider opening angles affects our results.

\subsection{Energy Scale and Event Rate} \label{sec:opening_angles_final}
The jet opening angle distribution affects the calculation of the burst true energy scales as well as the true event rate. The beaming-corrected energy is given by, $E_{\rm K}=f_b E_{\rm K,iso}$, where $f_b \equiv 1-cos(\theta_{\rm j})$ is the beaming correction factor and $E_{\rm K,iso}$ is the isotropic-equivalent kinetic energy. At the same time, emission from the relativistic jet is only immediately visible if the observer's line-of-sight is inside or intercepts the cone of the outflow. Thus, we expect the true event rate to be larger by a factor of $f_b^{-1}$, where $\mathfrak{R}_{\rm true} = f_b^{-1} \mathfrak{R}_{\rm obs}$.

For the SGRBs with opening angle measurements in our sample, we build the CDF of $f_b$ by calculating the beaming correction factors using every value that composes the distributions of opening angles (Section~\ref{sec:distribution_opening_angles}). We find a median of $\langle f_{b, {\rm det}} \rangle=0.6[-0.5,+3.0]\times 10^{-2}$ for the 10 bursts with well-measured opening angles. This value is consistent within errors to that quoted in \cite{Fong2015a}. The $f_{b}$ value is essentially a minimum value on the median beaming correction factor as the sample used to derive it comprises jet measurements which likely represent the narrowest jets. For the mock sample including wide jets, the median value expectedly shifts to a larger value of $\langle f_{b} \rangle=2.8[-2.6,+4.1]\times 10^{-2}$.

Using bursts with opening angle measurements, we find population median isotropic-equivalent kinetic and $\gamma$-ray energies of $\langle E_{\rm{K, iso}}\rangle \approx 3.7 \times 10^{51}$\,erg and $\langle E_{\gamma {\rm, iso}}\rangle \approx 9.7 \times 10^{50}$\,erg ($1-10000$\,keV), respectively, with a $\gamma$-ray efficiency of $\langle \eta_{\gamma} \rangle \approx 0.2$. Incorporating $\langle f_{b, {\rm det}} \rangle$, we obtain median beaming-corrected kinetic and $\gamma$-ray energies of $\langle E_{\rm K}\rangle \approx 2.1 \times 10^{49}$\,erg and $\langle E_{\rm{\gamma}}\rangle \approx 5.0 \times 10^{48}$\,erg, respectively. This results in a total beaming-corrected energy of $\langle E_{\rm{\rm true, tot}}\rangle \equiv \langle E_{\rm{K}}\rangle+\langle E_{\rm{\gamma}}\rangle \approx 2.6 \times 10^{49}$\,erg. If we assume similar isotropic-equivalent kinetic and $\gamma$-ray energies for the mock sample including wide jets, the median total beaming-corrected energy of the SGRB population increases to $\langle E_{\rm{\rm true, tot}}\rangle \approx 1.3 \times 10^{50}$\,erg. These values are similar to the energy scales found in previous works where $E_{\rm K} \sim 10^{49}$\,erg and $E_{\rm true} \sim 10^{50}$\,erg \citep{Berger2014a,Fong2015a,Jin2018}, respectively. In addition, even though the jet core of GRB\,170817A is more collimated than most of our SGRB measurements, its total inferred energetics are similar with $E_{\rm K} \sim 10^{49}-10^{51}$\,erg \citep{Margutti2021}.

In the case of event rates, we correct the observed local rate ($\mathfrak{R}_{\rm obs}$) derived from the SGRB luminosity function using the distribution of beaming factors. We note that several works which derive $\mathfrak{R}_{\rm obs}$ via the minimum $\gamma$-ray luminosity \citep[e.g.,][]{Guetta2006,Nakar2006,Coward2012,Wanderman2015,Ghirlanda2019,Liu2019}, have converged on $\mathfrak{R}_{\rm obs} \approx 5-10$~Gpc$^{-3}$~yr$^{-1}$. At the lowest end, \citet{Ghirlanda2019} derived $\mathfrak{R}_{\rm obs} \approx 0.5$\,Gpc$^{-3}$~yr$^{-1}$ based on available observational constraints of the {\it Fermi} SGRB population. However, this lower value is inferred from bright events with L$_{\rm{min, iso}} \approx 10^{51}$\,erg\,s$^{-1}$, ignoring a fraction of SGRBs that exhibit fainter luminosities. To account for the contribution of these fainter events with L$_{\rm {min, iso}}\approx 10^{49}$\,erg\,s$^{-1}$ \citep[e.g.,][]{Guetta2009,Wanderman2015}, we consider a local rate of $\mathfrak{R}_{\rm obs} \approx 10$\,Gpc$^{-3}$~yr$^{-1}$ \citep{Nakar2006} going forward.

\begin{figure}
\centering
\includegraphics[width=1.0\columnwidth]{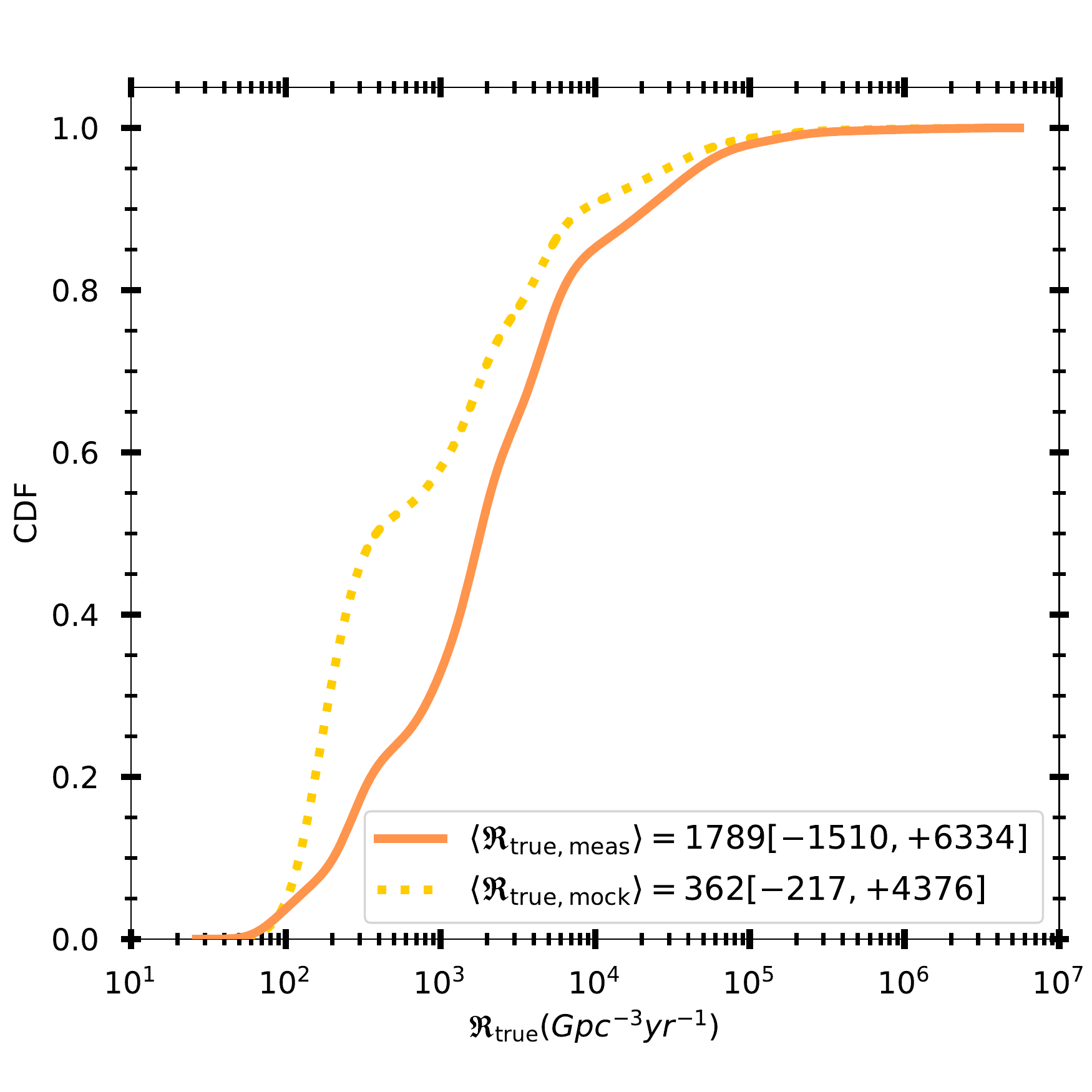}
\vspace{-0.1in}
\caption{The CDFs of true event rates for the opening angle measurement sample (orange solid line) and the mock sample including wide jets (yellow dashed line) assuming an observed local rate of $\mathfrak{R}_{\rm obs} \approx 10$\,Gpc$^{-3}$~yr$^{-1}$. Comparing both distributions, we see that a larger number of wider jet opening angles in our sample would lead to a more constrained true event rate median value.}
\label{fig:true_event_rate}
\end{figure}

Based on events with well-measured opening angles and $\mathfrak{R}_{\rm obs} \approx 10$\,Gpc$^{-3}$~yr$^{-1}$, we find a median value of $\langle \mathfrak{R}_{\rm true} \rangle = 1789[-1510,+6334]$~Gpc$^{-3}$~yr$^{-1}$ (Figure\,\ref{fig:true_event_rate}). This derived true event rate is consistent within errors to previously published values based on SGRBs, albeit is on the high end \citep[][]{Coward2012,Fong2014,Fong2015a,Jin2018,Dichiara2020} (Figure\,\ref{fig:true_event_rate}). This can be naturally explained because this rate comprises jet measurements which likely represent the narrower end of the population and ignores the existence of wider jets and lower limits. Thus, $\langle \mathfrak{R}_{\rm true} \rangle \approx 1790$\,Gpc$^{-3}$~yr$^{-1}$ can actually be considered a high estimate on the true event rate. If instead we use the SGRB mock sample which includes wider jets and is likely a better representation of the parent distribution, we find a lower value of $\langle \mathfrak{R}_{\rm true, mock} \rangle = 362[-217,+4376]$~Gpc$^{-3}$~yr$^{-1}$ (Figure\,\ref{fig:true_event_rate}). Therefore, one would expect the real SGRB event rate to lie between those derived from both the events with jet opening angle measurements and mock distribution.

%

\section{Discussion}\label{sec:discussion}
In this section, we discuss the implications of our findings in the context of the event energy scales and potential mechanisms to launch jets. We also compare the SGRB true event rate derived from our study with other observational and theoretical rates published in the literature for these events and BNS/NS-BH mergers. Additionally, we investigate what fraction of the SGRB population can be explained by these mergers. 

\subsection{Jet Opening Angles: Implications on Energy scales and Potential Launch Mechanism} \label{sec:Implications_energy}
The jet opening angles uniquely enable a determination of the true, beaming-corrected energy scale of SGRBs, which can be used to probe the energy extraction mechanism to power the jets. The calculation of the true energy for these events is key to discerning between the potential mechanisms to launch relativistic outflows \citep{Shibata2019}, either by neutrino pair ($\nu\bar{\nu}$) annihilation \citep{Jaroszynski1993,Mochkovitch1993,Rosswog2002} or as a magnetically driven jet \citep{Blandford1977,Rosswog2003,Rezzolla2011,Siegel2017} since one expects different released energy ranges. From our analysis, we find median beaming-corrected total energy releases of $E_{\rm true, tot} \approx (0.3-1) \times 10^{50}$\,erg.

In the $\nu\bar{\nu}$ annihilation scenario, launched jets have expected opening angles of $\theta_{\rm j}\sim 5-30^{\circ}$ and maximum beaming-corrected energies of $\sim10^{48-49}$\,erg \citep{Rosswog2002,Aloy2005,Birkl2007,Dessart2009,Murguia-Berthier2017}. \citet{Liu2015} and \citet{Perego2017} demonstrate that the deposited energy by the $\nu\bar{\nu}$ mechanism in comparison to the median inferred energy of $\approx 10^{50}$\,erg for detected SGRBs is not sufficient to power jets in these events. Even under the assumption of smaller opening angles of $\theta_{\rm j} < 10^{\circ}$, it is still not sufficient to explain SGRB jets powered by the energy extracted from the $\nu\bar{\nu}$ annihilation mechanism \citep{Perego2017}. Based on this, the $\nu\bar{\nu}$ annihilation mechanism is unlikely to be the dominant energy extraction mechanism to launch SGRBs.

 On the other hand, magnetohydrodynamic (MHD) processes \citep[i.e., Blandford-Znajek mechanism;][]{Blandford1977} can easily reach larger energy scales of $ \sim 10^{49-52}$\,erg \citep{Rosswog2005,Lee2007,Ruiz2016,Siegel2017}. We note that there are different predictions for the jet opening angles depending on the outflow's magnetization \citep{Rosswog2002,Duffell2018,Nathanail2020} with expected $\theta_{\rm j}\gtrsim 10^{\circ}$ for more magnetized jets \citep{Nathanail2020}. In addition, \citet{Christie2019} demonstrated with 3D MHD simulations that poloidal post-merger magnetic fields generate jets with $\theta_{\rm j}\sim 6^{\circ}-13^{\circ}$ and up to $E_{\rm K, iso}\sim 10^{52}$\,erg, while toroidal post-merger magnetic field geometry produces jets with $E_{\rm k, iso}\sim 10^{51}$\,erg and $\theta_{\rm j}\sim 3^{\circ}-5^{\circ}$. Indeed, this last magnetic field configuration of the post-merger disk is consistent with the jet opening angle and energetics found for GRB~170817A \citep{Margutti2021}, as well as the rather low efficiency of this event \citep[$\eta \sim 10^{-3}$;][]{Salafia2021}. Polarimetry studies specifically focused on the radio reverse shock of long GRBs have been essential to reveal their magnetic field configurations and hence the jet launching mechanism \citep[e.g.;][]{Granot2005,Laskar2019}. However, these type of studies are particularly difficult for SGRBs as they are fainter and evolve much faster than long GRBs so in the meantime, energetics can offer strong clues. Indeed, the agreement between the jet energetics and opening angles derived from the MHD scenarios and those inferred from our observations supports this scenario as the main mechanism to launch SGRB jets.

%
\begin{figure*}[!t]
\centering
\includegraphics[width=1.0\textwidth]{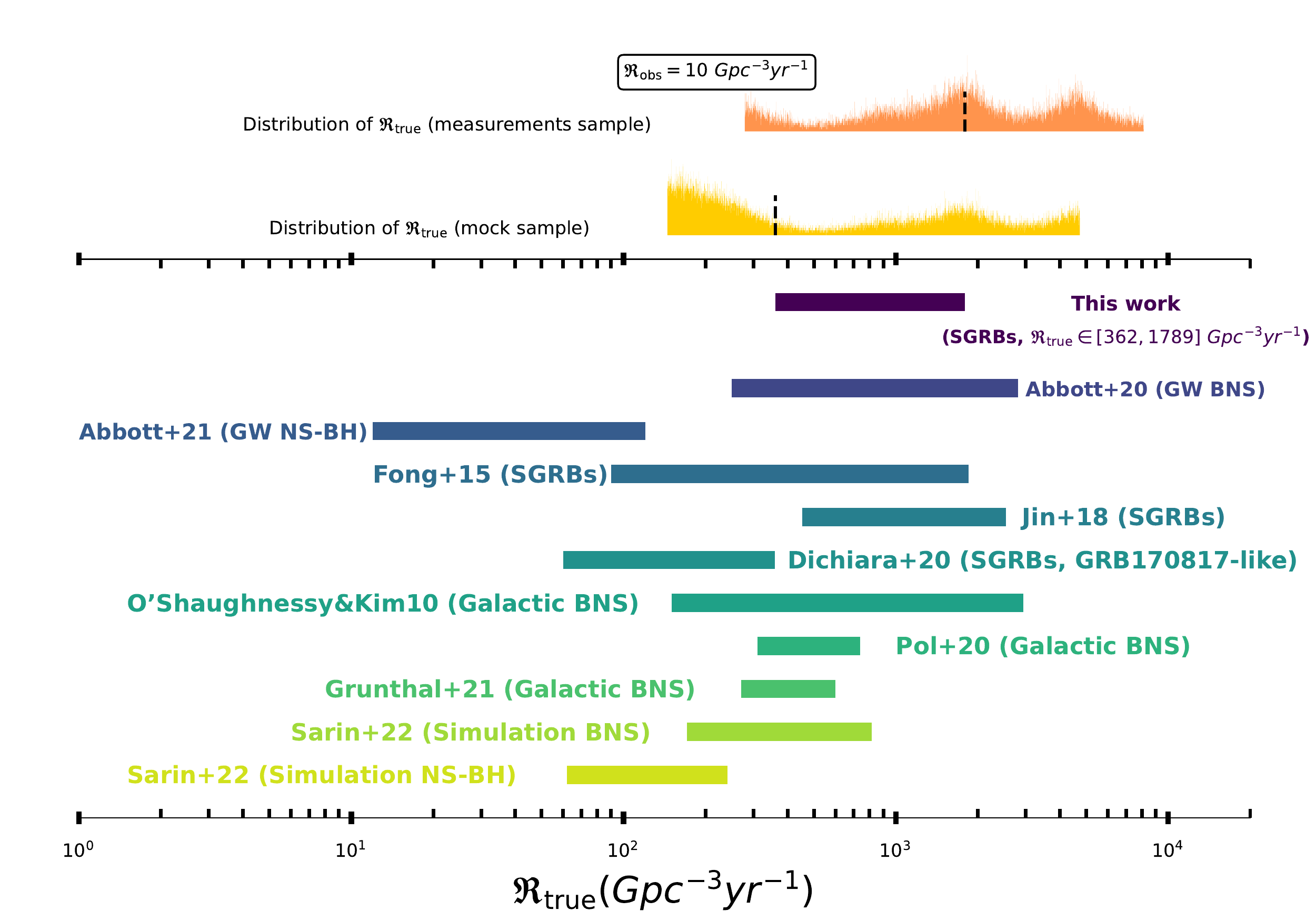}
\caption{\textit{upper panel:} The SGRB true event rate distributions ($\mathfrak{R}_{\rm true}$) derived from the opening angle measurements, $\langle \mathfrak{R}_{\rm true, meas} \rangle = 1789[-1510,+6334]$~Gpc$^{-3}$~yr$^{-1}$, (orange) and mock, $\langle \mathfrak{R}_{\rm true, mock} \rangle = 362[-217,+4376]$~Gpc$^{-3}$~yr$^{-1}$, (yellow) samples using the observed rate of $\mathfrak{R}_{\rm obs}=10$\,Gpc$^{-3}$~yr$^{-1}$. Median values of each distribution are indicated by black vertical lines. \textit{bottom panel:} We present a more realistic range of values of $\mathfrak{R}_{\rm true}$ that extends between the median values of the true event rate distributions derived from the opening angle measurements (vertical black dashed line) and mock samples (vertical grey dashed line). Other published rates are also shown as comparison. In particular, these rates are derived from: the detection of gravitational waves generated by BNS \citep{Abbott2020b} and NS-BH mergers \citep{Abbott2021}, SGRBs \citep{Fong2015a,Jin2018,Dichiara2020}, Galactic BNS \citep{OShaughnessy2010,Pol2020,Grunthal2021}, and the estimates for the BNS and NS-BH merger rates derived from population synthesis simulations by \citet{Sarin2022}. 68\% confidence levels are represented.}
\label{fig:True_Rate}
\end{figure*}
%

\subsection{Derived Event Rates and Implications for Compact Object Merger Progenitors}
One of the most important consequences of determining jet geometries is the inference on the true event rate ($\mathfrak{R}_{\rm true}$). In particular, the estimate of $\mathfrak{R}_{\rm true}$ for SGRBs can be compared with the rates of BNS ($\mathfrak{R}_{\rm BNS}$) or NS-BH ($\mathfrak{R}_{\rm NS-BH}$) mergers derived from the Advanced LIGO/Virgo gravitational wave detections \citep{Abbott2017a,Abbott2017b,Abbott2017c,Abbott2020a,Abbott2021}. This provides us with basic information on the fractions of BNS and NS-BH mergers that may power SGRBs and launch jets. Coupled with ejecta masses inferred from kilonovae \citep{Gompertz2018,Rossi2020,Rastinejad2021}, the true event rate can also determine the role of BNS/NS-BH mergers in populating the universe with $r$-process elements (e.g., \citealt{Kasen2017,Hotokezaka2018, Rosswog2018}

To date, several works have derived an observed event rate of $\mathfrak{R}_{\rm obs} \approx 5-10$~Gpc$^{-3}$~yr$^{-1}$ \citep[e.g.,][]{Guetta2006,Nakar2006,Coward2012,Wanderman2015,Ghirlanda2019,Liu2019} using the minimum $\gamma$-ray luminosity of SGRBs. However, the lower end of $\mathfrak{R}_{\rm obs} \approx 5$~Gpc$^{-3}$~yr$^{-1}$ is derived from the most luminous events. Therefore, for calculating the SGRB true event rate from the jet opening angle measurements and mock distributions, we choose the value of $\mathfrak{R}_{\rm obs} \approx 10$\,Gpc$^{-3}$~yr$^{-1}$ \citep[assuming L$_{\rm {iso, min}} \approx 10^{49}$\,erg\,s$^{-1};$][]{Nakar2006}, which has been used in previous literature \citep[][]{Metzger2012,Fong2015a,Mandel2022}. In Figure\,\ref{fig:True_Rate}, we present our distributions of $\mathfrak{R}_{\rm true}$ and compare these rates with BNS and NS-BH merger rates published in the literature as well as previous SGRB studies. We use the range of rates, $R_{\rm true} \approx 360-1790$~Gpc$^{-3}$~yr$^{-1}$, where the lower end is set by the mock sample including wider jets, while the upper end is set by jet measurements only. This range is fully consistent with the estimated range of Galactic BNS merger rates \citep{OShaughnessy2010,Pol2020,Grunthal2021} and the BNS merger rate derived from the detection of gravitational waves \citep[Figure\,\ref{fig:True_Rate};][]{Abbott2020a}. This implies that SGRBs are predominantly the result of BNS mergers with successfully launched relativistic jets, with the most successful direct evidence being the joint detection of the BNS merger GW170817 with its SGRB \citep[GRB~170817A;][]{Abbott2017a}.

An open question remains on the fraction of SGRBs which are derived from NS-BH mergers. Thus far, the identification of EM counterparts to the few known GW-detected NS-BH mergers, GW200105 and GW200115 \citep[e.g.,][]{Anand2021,Dichiara2021,Rastinejad2021} have not yielded EM counterparts. Theoretically, one expects the partial or complete disruption of the NS by the BH, resulting in little or no ejected matter and electromagnetic emission \citep[e.g.,][]{Foucart2012}. However, there are conditions for which NS-BH mergers can produce successful SGRBs and/or EM emission in simulations \citep[e.g.,][]{Bhattacharya2019,Barbieri2020,Darbha2021}. Observational studies have also hinted at the possibility of a NS-BH merger contribution from the population of SGRBs with extended emission \citep{Gompertz2020}. However, from Figure\,\ref{fig:True_Rate}, our SGRB range of rates are $\gtrsim 13$ times larger than the rates of NS-BH mergers inferred from both GW observations \citep{Abbott2021} and population synthesis simulations \citep{Sarin2022}. Therefore, our results only support that (at most) a small fraction of SGRBs originate from NS-BH mergers (Figure\,\ref{fig:True_Rate}).

In our study, we also uncover a group of SGRBs with inferred jet opening angles of $\theta_{\rm j}\gtrsim 10^{\circ}$, which are broader than the derived median of jet opening angle measurements of $\langle \theta_{\rm j} \rangle \approx 6^{\circ}$. For most of these cases, the lower limits on the opening angles were inferred thanks to late-time X-ray detections obtained at $t_{\rm j}>10$\,days. As seen in our study, the inclusion of even a few wide jets pushes the inferred true event rates to lower values, which has consequences for the progenitors of SGRBs. Additionally, successful jets from NS-BH mergers are expected to be wider ($\theta_{\rm j} \approx 25-30^{\circ}$), due to the lower densities of the surrounding environment. In addition, the widest jets are expected to originate from mergers containing highly spinning BHs \citep[][]{Murguia-Berthier2017,Ruiz2018}. However, jets wider than $\approx 30^{\circ}$ are not expected to routinely survive, resulting in a failed or choked jet \citep{Ghirlanda2019}. Thus, continued X-ray monitoring of cosmological SGRBs to late times, in tandem with monitoring of future GW-detected SGRBs events, will be imperative in constraining the true population of wide jets.

\section{Conclusions}\label{sec:conclusions}
We have presented a comprehensive compilation of {\it Swift} SGRBs discovered between 2005 and 2021, and observed with \textit{Chandra} and \textit{XMM-Newton} at late times ($\delta t > 0.8$\,days). We conclude our findings below:
\begin{itemize}
    \item From the 29 SGRBs in our final sample: 18 were observed by \textit{Chandra}, 4 by \textit{XMM-Newton} and 7 by both observatories, resulting in 60 epochs, that we uniformily-analyzed, across all events at $\delta t > 0.8$\,days.
    \item Using broadband information and applying the synchrotron afterglow model, we find $n_0 \approx 0.7\times10^{-5}-0.8$\,cm$^{-3}$ and $E_{\rm K,iso} \approx 3\times10^{49}-9\times10^{53}$\,erg, and median values of $\langle n_0 \rangle \approx 4.1 \times 10^{-4}$\,cm$^{-3}$ and $\langle E_{\rm K,iso} \rangle \approx 3.7 \times 10^{51}$\,erg.
    \item We identify nine SGRBs (SGRBs~050724A, 051221A, 111020A, 130603B, 140903A, 150424A, 160821B, 200411A and 200522A) with significant steepenings in their X-ray light curves that are best explained by jet breaks, two of which are new identifications (SGRBs~050724A and 200411A). Including the wide-angle GRB~211106A with an identified break in its radio afterglow light curve, we find a range of jet break times between $\approx 0.1-30$ days after the bursts, translating to $\langle \theta_{\rm j,det} \rangle=6.1^{\circ}[-3.2^{\circ},+9.3^{\circ}]$ (68\% confidence on the entire distribution).
    \item From the non-detection of jet breaks for 19 events, we derive lower limits on the opening angles of $\theta_{\rm j}\gtrsim 0.3-26^{\circ}$, including 12 new limits. Of particular interest are six events with wide inferred opening angles of $\theta_{\rm j}\gtrsim 10^{\circ}$. Coupled with two wide-angle events, SGRBs\,050724A and 211106A with $\theta_{\rm j}\approx 34^{\circ}$ and $\approx 16^{\circ}$, respectively, we have unveiled a growing population of SGRBs with wide jets.
    \item We obtain beaming-corrected total true energies between $E_{\rm true,tot}\approx 10^{49}-10^{50}$, which are consistent with MHD processes as the mechanism of energy extraction to launch jets.
    \item We derive a range for beaming-corrected true event rates of $\mathfrak{R}_{\rm true} \approx 360-1790$~Gpc$^{-3}$~yr$^{-1}$, for which the low end is set by the inclusion of wider jets, and the upper end is set by including jet measurements alone. These rates are fully consistent with the rates of BNS mergers derived from GW events, as well as the rates of Galactic BNS mergers. This aligns with expectations that the predominant progenitor channel of SGRBs is BNS mergers. It is also plausible, although cannot be confirmed given current rate uncertainties, that most BNS mergers produce successful SGRB jets.
    \item The SGRB event rate is $\gtrsim 13$~times larger than the GW-derived NS-BH merger rate. Thus, we find that (at most) a small fraction of SGRBs could originate from these mergers.
\end{itemize}

\noindent Our study highlights the importance of the late-time X-ray monitoring of SGRBs in constraining the beaming angles of SGRBs. SGRBs not only exhibit narrow opening angles of $\approx 6^{\circ}$, but are also capable of launching broader jets with opening angles of $\gtrsim 10^{\circ}$. In recent years, a concerted effort has been made to follow-up SGRB X-ray afterglows beyond 1 day after the burst trigger, when they are expected to be fainter and exhibit breaks. This has resulted in a substantial increase in the number of jet opening angle measurements and meaningful lower limits. Additionally, the exceptional coordination between the most sensitive space- and ground-based telescopes has allowed us to perform broadband monitoring of these events, providing constraints on the basics of bursts energetics and environmental properties. Relative to {\it Swift}/XRT, the exceptional sensitivity of \textit{Chandra} and \textit{XMM-Newton} has enabled monitoring up to $60$~days after the burst in some cases. Moreover, the high spatial resolution of \textit{Chandra} enables us to disentangle the afterglow from any contaminating X-ray source. The combined capabilities of these observatories have enabled studies to confront different jet launching mechanisms and further understand their neutron star merger progenitors. The next generation of X-ray missions like NewAthena and XRISM will be indispensable to maintain sensitivity to late jet breaks, and thus wider jets. 

Our study provides a baseline for the geometries of successfully-launched jets from mergers. The first and only joint detection of GW170817/SGRB\,170817A (170817A) with a successfully launched jet  \citep[][]{Abbott2017a,Goldstein2017,Savchenko2017} enabled a tight constraint on the opening angle of $\approx 2-4^{\circ}$, and evidence for jet structure \citep[e.g.][]{Lamb2017,Alexander2018,DAvanzo2018,Margutti2018,Troja2018,Xie2018,Fong2019,Hajela2019}. In tandem with the past two decades of SGRB observations, the upcoming observing run (O4) of the Advanced LIGO/Virgo/KAGRA \citep{Abbott2020b} and beyond will undoubtedly provide a complementary view on the progenitor conditions necessary to launch jets.

\section{Acknowledgements}
The authors acknowledge Phil Evans for the extensive and constructive discussion on how to obtain the full set of unabsorbed X-ray flux light curves for the GRB afterglows from the UK Swift Science Data Centre. The authors also thank Amy Lien for her valuable exchange of ideas on calculating the isotropic $\gamma$-ray energy of the events.

The Fong Group at Northwestern acknowledges support by the National Science Foundation under grant Nos. AST-1814782, AST-1909358 and CAREER grant No. AST-2047919. W.F. gratefully acknowledges support by the David and Lucile Packard Foundation, the Alfred P. Sloan Foundation, and the Research Corporation for Science Advancement through Cottrell Scholar Award \#28284. The Berger Time Domain Group at Harvard is supported by NSF and NASA grants. TL acknowledges support from the Radboud Excellence Initiative. RM acknowledges support from the NSF (grant numbers AST-2224255 and AST-2221789). BDM acknowledges support from the NSF (grant number AST-2002577).

Support for this work was provided by the National Aeronautics and Space Administration through Chandra Award Numbers GO0-21041X, GO0-21042X, and GO1-22043X issued by the Chandra X-ray Center, which is operated by the Smithsonian Astrophysical Observatory for and on behalf of the National Aeronautics Space Administration under contract NAS8-03060. The scientific results reported in this article are based to a significant degree on observations made by the \textit{Chandra} X-ray Observatory, and data obtained from the \textit{Chandra} Data Archive. In addition, this research has made use of software provided by the \textit{Chandra} X-ray Center (CXC) in the application package \texttt{CIAO}. This work is based on observations obtained with \textit{XMM-Newton}, an ESA science mission with instruments and contributions directly funded by ESA Member States and NASA. This work made use of data supplied by the UK Swift Science Data Centre at the University of Leicester.

\vspace{5mm}
\facilities{\textit{FERMI}/GBM, \textit{Swift}(BAT and XRT), \textit{Chandra}(ACIS-S and HRC), \textit{XMM-Newton}(EPIC)}

\software{\texttt{CIAO} software package \citep[v.4.12][]{Fruscione2006}, \texttt{emcee} package \citep{Foreman-Mackey2013,Foreman-Mackey2019}, \texttt{HEASoft} software \citep[v.6.17;][]{Blackburn1999,NASA2014}, \texttt{NumPy} \citep{vanderWalt2011}, \texttt{Pandas} \citep{Mckinney2010}, \texttt{Matplotlib} \citep{Hunter2007}, \texttt{SAS} software \citep[v.18.0.0;][]{Gabriel2004}, \texttt{XSPEC} \citep{Arnaud1996}}

\clearpage

\begin{appendix}
\label{sec:appendix}
\section{Fitting Models for the X-ray Afterglow Light Curves} \label{sec:X-ray_LC_models}
The following lines show the models used to fit the X-ray afterglow light curves in Section\,\ref{sec:fitting_method_LC}:
\begin{itemize}
  \item Single power-law model
\end{itemize}
\begin{equation}
\large
    F_{\rm X} = C t^{\alpha_{1}}
\label{eqn:SPL}
\end{equation}

\begin{itemize}
  \item Broken power-law model
\end{itemize}
\begin{equation}
\large
    F_{\rm X} = C  \left[ \left[  \frac{ t } {t_{b_{1}} }  \right]^{-s_{1}\alpha_{1}} +  \left[  \frac{ t } {t_{b_{1}} }  \right]^{-s_{1}\alpha_{2}} \right]^{-1/s_{1}}
\label{eqn:BPL}
\end{equation}

\begin{itemize}
  \item Triple power-law model
\end{itemize}
\begin{equation}
\large
    F_{\rm X} = C  \left[\left[ \left[  \frac{ t } {t_{b_{1}} }  \right]^{-s_{1}\alpha_{1}} + \left[  \frac{ t } {t_{b_{1}} }  \right]^{-s_{1}\alpha_{2}} \right]^{-1/s_{1}} \times \left[ 1+ \left[ \frac{ t } {t_{b_{2}}} \right]^{-s_{2}(\alpha_{2}-\alpha_{3})}\right]^{-1/s_{2}} \right]
\label{eqn:TPL} 
\end{equation}

\noindent where F$_{\rm X}$ is the X-ray unabsorbed flux (erg~s$^{-1}$~cm$^{-2}$) of the afterglow, $C$ is the normalization or amplitude, $\alpha_{i}$ ($i=1,2,3$) correspond to the temporal decay index, $t_{bj}$ ($j=1,2$) are the break times (in seconds) and $s_{j}$ are the constant smoothness parameters ($s=-10$ or $s=10$, depending if the change in slope between power-law segments is positive or negative, respectively). We note that the break times are not necessarily classified as jet breaks at this point in the process.

\section{F-test chart flow} \label{sec:F_test}
\begin{figure*}[!h]
    \centering
    \tikzstyle{none} = [draw=none, fill=none, text width=2em, text centered, minimum height=1em]
    \tikzstyle{block1} = [rectangle, draw, fill=blue!20, text width=23em, text centered, rounded corners, minimum height=2.5em]
    \tikzstyle{block2} = [rectangle, draw, fill=yellow!20, text width=20em, text centered, rounded corners, minimum height=2em]
    \tikzstyle{block3} = [rectangle, draw, fill=cyan!20, text width=6em, text centered, rounded corners, minimum height=2em, node distance=3cm]
    \tikzstyle{block4} = [rectangle, draw, fill=blue!20, text width=24em, text centered, rounded corners, minimum height=2.5em, node distance=3.5cm]
    \tikzstyle{line} = [draw, -latex']
    \tikzstyle{cloud} = [draw, ellipse,fill=red!20, node distance=2cm, minimum height=0.5em]

    \begin{tikzpicture}[node distance = 1.5cm, auto]  \label{chart:flowchart}
        \node [block1] (null-hypothesis) {$H_0$: Single power law (SPL) is the best-fit model \\
        $H_1$: we need a more complex model};
        \node [block2, below of=null-hypothesis] (Eq1) {$F_{{\rm stat},1}(\chi^2_{1,{\rm SPL}}, \chi^2_{2,{\rm BPL}}, k_{1,{\rm SPL}}, k_{2,{\rm BPL}})$};
        \node [none, below of=Eq1](juntion){\textit{if}};
        \node [block3, left of=juntion] (ftest1) {$F_{{\rm stat},1}<F_{\rm crit}$};
        \node [cloud, below of=ftest1] (H0) {$H_0$ Accepted};
        \node [block3, right of=juntion] (ftest1') {$F_{{\rm stat},1}>F_{\rm crit}$};
        \node [block4, below of=ftest1'] (null-hypothesis2) {$H'_0$: Broken power law (BPL) is the best-fit model \\
        $H'_1$: we need triple power law (TPL) model};
        \node [block2, below of=null-hypothesis2] (Eq2) {$F_{{\rm stat},2}(\chi^2_{1,{\rm BPL}}, \chi^2_{2,{\rm TPL}}, k_{1,{\rm BPL}}, k_{2,{\rm TPL}})$};
        \node [none, below of=Eq2](juntion2){\textit{if}};
        \node [block3, left of=juntion2] (ftest2) {$F_{{\rm stat},2}<F_{\rm crit}$};
        \node [block3, right of=juntion2] (ftest2') {$F_{{\rm stat},2}>F_{\rm crit}$};
        \node [cloud, below of=ftest2] (H0') {$H'_0$ Accepted};
        \node [cloud, below of=ftest2'] (H1') {$H'_1$ Accepted};
        \path [line] (null-hypothesis) -- (Eq1);
        \path [line] (Eq1) -- (juntion);
        \path [line] (juntion) -- (ftest1);
        \path [line] (ftest1) -- (H0);
        \path [line] (juntion) -- (ftest1');
        \path [line] (ftest1') -- (null-hypothesis2);
        \path [line] (null-hypothesis2) -- (Eq2);
        \path [line] (Eq2) -- (juntion2);
        \path [line] (juntion2) -- (ftest2);
        \path [line] (juntion2) -- (ftest2');
        \path [line] (ftest2) -- (H0');
        \path [line] (ftest2') -- (H1');
    \end{tikzpicture}
\caption{The F-test flow chart followed to the best-fit model of the SGRB X-ray afterglow light curves.}
\label{fig:flowchart}
\end{figure*}
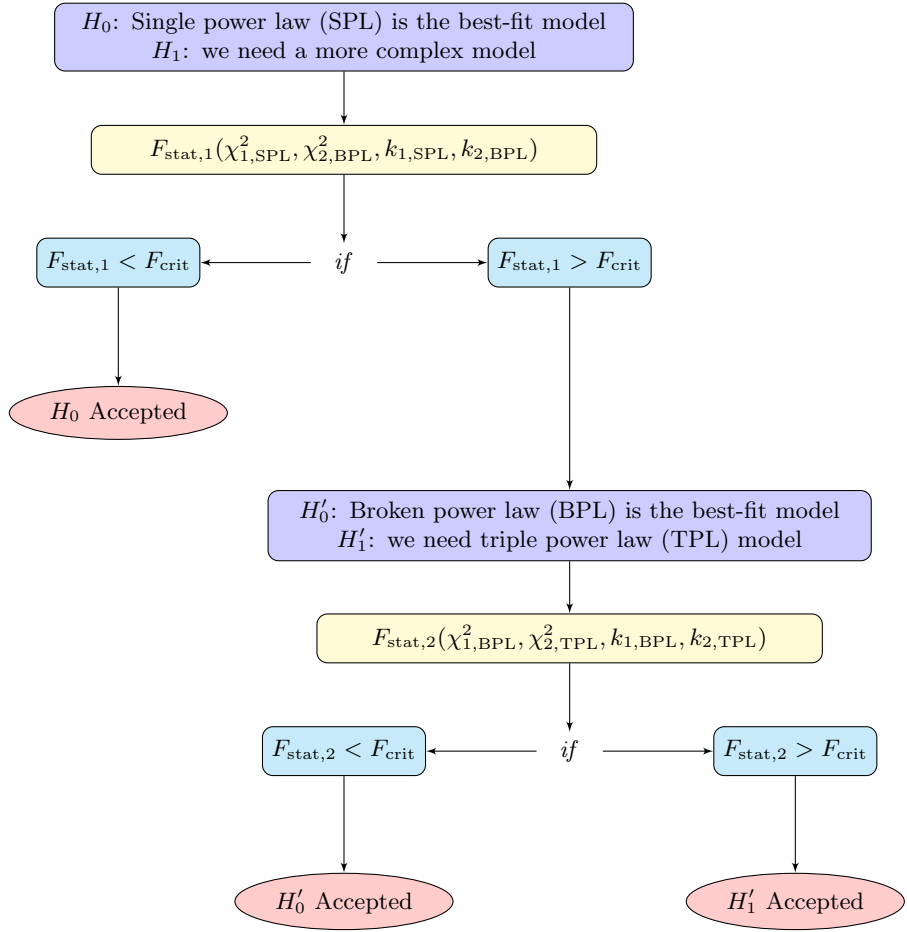

In Section\,\ref{sec:fitting_method_LC}, we apply an F-test to discern which is the best-fit model between the three different fitting models used in this work (see Appendix\,\ref{sec:X-ray_LC_models}). Thus, we establish an F-test null-hypothesis ($H_0$) in which we accept that a simpler model is better to describe the data. Treating the $\chi^{2}$ values from each fit as random variables that follow an F-distribution, we can define an F-statistic value assuming the ratio between them:

\begin{equation}
\Large
    F_{\rm stat} = \frac{\frac{(\chi^{2}_{1}-\chi^{2}_{2})}{(k_2-k_1)}}{\frac{\chi^{2}_{2}}{(n-k_2)}}
    \label{eqn:Ftest}
\end{equation}

\noindent where $\chi^{2}_{1}$ is the $\chi^{2}$ value of the simpler model and $\chi^{2}_{2}$ corresponds to the $\chi^{2}$ value of the more complex model, $k_1$ and $k_2$ are the number of variables in the simpler and more complex models, respectively, and $n$ is the number of data points. We also compute a critical value ($F_{\rm crit}$) with a confidence interval of $95\%$ ($\alpha = 0.05$) and compare it against $F_{\rm stat}$ to accept or reject $H_0$. If $F_{\rm stat}>F_{\rm crit}$, we then reject $H_0$ and accept the alternate hypothesis ($H_1$), i.e., we need a more complex model to describe the SGRB light curve. Since we have three different models (simple, broken and triple power-law models), we follow the flow chart in Figure\,\ref{fig:flowchart} to determine which of those models describes better the SGRBs light curves.

%
\vspace{0.1in}
%

\section{X-ray observations of GRB~130603B} \label{sec:panel_GRB130603B}

%
\begin{figure}[H]
\centering
\includegraphics[width=1.0\textwidth]{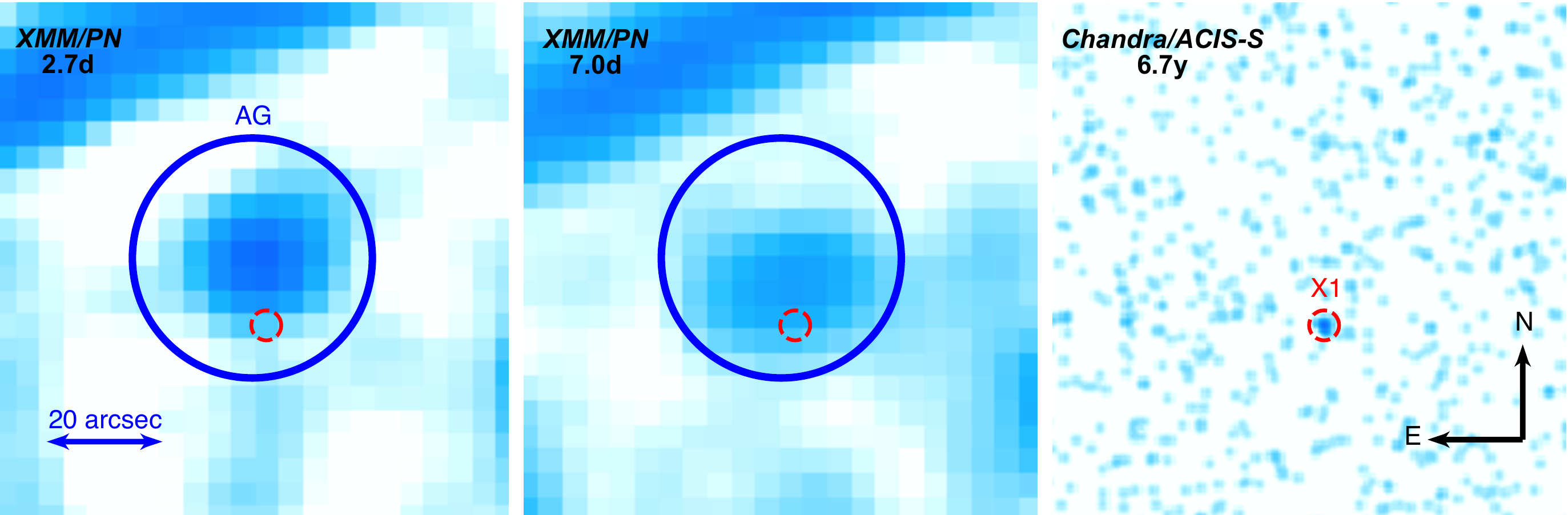}
\caption{X-ray imaging panel of GRB~130603B. The \textit{XMM-Newton}/PN observations (left and middle) at $\delta t~\sim2.7$ and $\sim7.0$~days in the $0.3-10$\,keV energy band, and merged \textit{Chandra}/ACIS-S observation (right) obtained at $\delta t \sim 6.7$~years in the $0.5-7$\,keV energy band. The blue circle indicates the \textit{XMM-Newton} source region. The small dashed red region in the images shows the contaminating source (X1), which is resolved in the \textit{Chandra} observation. The final X-ray flux afterglow light curve is corrected against the extra contribution of this contaminant, which was not taken into account in earlier works.}
\label{fig:panel_GRB130603B}
\end{figure}
%

\end{appendix}

\clearpage

\bibliography{reference}{}
\bibliographystyle{aasjournal}

\end{document}